\documentclass[
prd,aps,amsmath,superscriptaddress,nofootinbib,
]{revtex4}

\usepackage[hypertex]{hyperref}
\usepackage{graphicx}
\usepackage{epstopdf}
\usepackage{bm}
\usepackage{epsfig}
\usepackage{xspace}
\usepackage{color}

  \newcount\hour \newcount\minute
  \hour=\time \divide \hour by 60
  \minute=\time
  \newcommand{\mydate}{\ \today \ - \number\hour :\ifnum \minute<10 0\fi 
\number\minute}

\def\OMIT#1{}

\newcommand{\nn}{\nonumber}

\newcommand{\bea}{\begin{eqnarray}}
\newcommand{\eea}{\end{eqnarray}}

\def\lsim{\mathrel{\raise.3ex\hbox{$<$\kern-.75em\lower1ex\hbox{$\sim$}}}}
\def\gsim{\mathrel{\raise.3ex\hbox{$>$\kern-.75em\lower1ex\hbox{$\sim$}}}}

\begin{document}
\setlength\baselineskip{12pt}



\title{Heavy Quark Vacuum Polarization Function at ${\cal O}(\alpha_s^2)$ 
and ${\cal O}(\alpha_s^3)$\\[3mm]\mbox{}} 

\vspace*{1cm}

\author{Andr\'e H.~Hoang\footnote{Electronic address: ahoang@mppmu.mpg.de}}
  \affiliation{Max-Planck-Institut f\"ur Physik\\ (Werner-Heisenberg-Institut)\\
  F\"ohringer Ring 6,  80805 M\"unchen, Germany\\[5mm]}


\author{Vicent Mateu\footnote{Electronic address:  mateu@mppmu.mpg.de}}
  \affiliation{Max-Planck-Institut f\"ur Physik\\ (Werner-Heisenberg-Institut)\\
  F\"ohringer Ring 6,  80805 M\"unchen, Germany\\[5mm]}

\author{S.~Mohammad Zebarjad\footnote{Electronic address: zebarjad@susc.ac.ir}}
  \affiliation{Max-Planck-Institut f\"ur Physik\\ (Werner-Heisenberg-Institut)\\
  F\"ohringer Ring 6,  80805 M\"unchen, Germany\\[5mm]}
  \affiliation{Physics Department, College of Sciences, Shiraz University\\
  Shiraz 71454, Iran}



\begin{abstract}
\vspace{0.4cm}
\setlength\baselineskip{12pt}
We determine the full mass and $q^2$ dependence of the heavy quark vacuum
polarization function $\Pi(q^2)$ and its contribution to the total
$e^+e^-$ cross section at ${\cal O}(\alpha_s^2)$ and 
${\cal O}(\alpha_s^3)$ in perturbative QCD. We use known results for the
expansions of $\Pi(q^2)$ at high energies, in the threshold region and around
$q^2=0$, conformal mapping and the Pad\'e approximation method.
From our results for $\Pi(q^2)$ we determine numerically at ${\cal
  O}(\alpha_s^3)$ the previously unknown non-logarithmic contributions in the  
high-energy expansion at order $(m^2/q^2)^i$ for $i=0,1$ and the coefficients
in the expansion around $q^2=0$ at order $q^{2n}$ with $n\ge 2$. 
We also determine at ${\cal O}(\alpha_s^2)$ the previously unknown 
${\cal O}(v^0)$ constant term in the expansion of $\Pi(q^2)$  
in the threshold region, where $v$ is the quark velocity. 
Our method allows for a quantitative estimate of uncertainties and can
be systematically improved once more information in the three 
kinematic regions becomes available by future multi-loop computations.
For the contributions to the total $e^+e^-$ cross section at ${\cal
  O}(\alpha_s^2)$ we confirm results obtained earlier by Chetyrkin, K\"uhn and
Steinhauser. 
\\[25pt]
 \hbox{MPP-2008-83}\\
 \hbox{arXiv:0807.4173          [hep-ph]}
\end{abstract}

\maketitle

%
%
%
%

\newpage
\section{Introduction}
\label{sectionintroduction}

The vacuum polarization function $\Pi(q^2)$ defined by the correlator of two
electromagnetic currents $j^\mu(x)=\bar\psi(x)\gamma^\mu\psi(x)$,
\begin{align}
\label{pidef}
\left(g_{\mu\nu}q^2-q_\mu q_\nu\right)\, \Pi(q^2) 
\, = \, \, 
- \,i
\int\mathrm{d}x\, e^{iqx}\left\langle \,0\left|T\, j_\mu(x)j_\nu(0)\right|0\,
\right\rangle
\,,
\end{align}
where $q^\mu$ is the four-momentum of the quark pair produced or annihilated
by $j^\mu$, represents an important quantity for theoretical studies as well
as for many practical phenomenological applications. Relevant applications for
the case of massive quarks include predictions of the hadronic cross section
$R\sim \mbox{Im}[\Pi]$, or sum rules for the determination of the heavy quark
masses~\cite{Novikov:1977dq,Reinders:1984sr}. These sum rules are based on
moments of the cross section for heavy 
quark pair production 
\begin{align}
M_n \, = \int_{4m^2}^\infty\dfrac{\mathrm{d}s}{s^{n+1}}\,R(s)
\,,
\end{align}
which in fixed-order perturbation theory are related to the expansion
coefficients of $\Pi(q^2)$ around $q^2= 0$,
\begin{align}
\Pi(q^2\approx 0,m^2) \, = \, \dfrac{1}{12\,\pi^2\,Q_q^2}\sum_{n=1}^\infty 
M_n\, q^{2n}
\,,
\end{align}
where $Q_q$ is the heavy quark electric charge.
In general the knowledge of the full dependence of the vacuum
polarization function $\Pi(q^2)$ on $q^2$ and the quark mass $m$ is desirable
to avoid having to rely on approximations that are only valid in certain
kinematic regimes.  

At ${\cal O}(\alpha_s)$ the full mass and $q^2$ dependence of the vacuum
polarization function is known from analytic computations carried out in
Ref.~\cite{Kallen:1955fb}. At 
${\cal O}(\alpha_s^2)$  analogous analytic results exist for the contributions
that originate from inserting the massive~\cite{Kniehl:1989kz,Hoang:1994it}
and massless~\cite{Hoang:1995ex} 
fermion loops into ${\cal O}(\alpha_s)$ one-gluon exchange diagrams. 
For the other ${\cal O}(\alpha_s^2)$ contributions results for the expansions of
$\Pi(q^2)$ in the high-energy limit, $|q^2|\to\infty$, the nonrelativistic
threshold regime, $q^2\approx 4m^2$, and in the Euclidean region around
$q^2=0$ were used to reconstruct an accurate
approximation~\cite{Chetyrkin:1995ii,Chetyrkin:1996cf}. The method is based on 
the definition 
of subtraction functions which account for all logarithmic terms that arise
for the expansions in the high-energy limit and in the threshold
region. Using a conformal transformation to a new variable $\omega$ the full
$q^2$ and mass dependence in the complex plane of the remaining contributions
can be mapped into the unit circle rendering those contributions to an
analytic function in the variable $\omega$. The latter can then be successfully
approximated by Pad\'e approximants using the remaining expansion coefficients
that are not related to logarithmic terms. With a large number of expansion
coefficients for the three kinematic limits the full mass and $q^2$ dependence
of the ${\cal O}(\alpha_s^2)$ vacuum polarization function can be determined with 
small numerical uncertainties.  

For the ${\cal O}(\alpha_s^3)$ vacuum polarization there is also no fully
analytic result available in the literature. 
A numerical study of the full ${\cal O}(\alpha_s^3 n_f^2)$ double fermionic
contributions to the vacuum polarization function can be found in
Ref.~\cite{Czakon:2007qi}.   
In the high-energy expansion its contributions to the total cross section up to order
$(m^2/q^2)^2$ are known~\cite{Chetyrkin:2000zk}. A comprehensive review of
these results can be found in Ref.~\cite{Harlander:2002ur}. Moreover, in the
threshold region, where an expansion in the small quark velocity $v$ can be
carried out, the  
${\cal O}(\alpha_s^3)$ contributions to the total cross section at order
$1/v^i$ for $i=2,1,0$ are available from a factorization theorem for the heavy
quark-antiquark pair 
production cross section in nonrelativistic QCD (NRQCD) at
next-to-next-to-leading order (NNLO)~\cite{Hoang:1997sj,Hoang:1998xf}. More
recently also the moments  
$M_1$~\cite{Chetyrkin:2006xg,Boughezal:2006px} and 
$M_2$~\cite{Maier:2008he} at ${\cal O}(\alpha_s^3)$ have become 
available using elaborate high-power computer algebra tools.

In this work we use the presently available information on the 
${\cal O}(\alpha_s^3)$ corrections to the vacuum polarization function
$\Pi(q^2)$ in the high-energy limit, in the threshold region and the small
$q^2$ domain to reconstruct the full $q^2$ and mass dependence of the vacuum
polarization function at ${\cal O}(\alpha_s^3)$. 
The method we use is similar to the approach of
Refs.~\cite{Chetyrkin:1995ii,Chetyrkin:1996cf} employed 
previously for the ${\cal O}(\alpha_s^2)$ corrections of the vacuum
polarization function (see also
Refs.~\cite{Baikov:1995ui,Fleischer:1994ef,Broadhurst:1993mw}), but also
accommodates a few 
notable differences which are motivated by the fact that less information
is known on the vacuum polarization function at ${\cal O}(\alpha_s^3)$. 
While in Refs.~\cite{Chetyrkin:1995ii,Chetyrkin:1996cf}
information from the high-energy expansion up to order $(m^2/q^2)$ and from
the threshold expansion up to next-to-leading order (NLO) was
incorporated for the reconstruction, we account for the expansions up to order
$(m^2/q^2)^2$ at high energies and up to NNLO in the threshold region.
While in Ref.~\cite{Chetyrkin:1995ii,Chetyrkin:1996cf} the full set  
of terms in the high-energy expansion of $\Pi$ was used for the construction,
in this work we only rely on the terms that carry an absorptive part and that 
contribute to the cross section above threshold.
We show that our method allows
to determine previously unknown non-logarithmic terms of the vacuum
polarization at ${\cal O}(\alpha_s^3)$ in the high-energy expansion with very
small errors. Moreover, while
in Ref.~\cite{Chetyrkin:1995ii,Chetyrkin:1996cf} and later in
Ref.~\cite{Chetyrkin:1997mb} the coefficients of the small-$q^2$ expansion
were included  up to order $q^{14}$ and $q^{16}$, respectively, we only rely
on the presently available 
${\cal O}(\alpha_s^3)$ coefficients up to order $q^4$. Our method allows to
determine 
the expansion coefficients at order $q^{2n}$ with $n\ge 3$. The results allow
to compute the corresponding moments $M_n$ in the fixed-order
expansion at ${\cal O}(\alpha_s^3)$. For phenomenologically relevant values of
$n$ the error in the $M_n$ due to the uncertainties in these coefficients is
an order of magnitude smaller than the remaining scale-uncertainties of the 
$M_n$ at ${\cal O}(\alpha_s^3)$ . 
We demonstrate the reliability of the results by using the same approach for
determining the corresponding coefficients for the vacuum polarization
function at ${\cal O}(\alpha_s^2)$ where
their values are well known analytically from the computations of Feynman
diagrams. Another noteworthy difference of our approach to
Refs.~\cite{Chetyrkin:1995ii,Chetyrkin:1996cf} is that we implement a
continuous set of subtraction functions to have a more reliable estimation of
the uncertainty inherent to the method.
Our approach can systematically incorporate new information from
the expansions in the three kinematical regions, once it becomes available.

One important application of the vacuum polarization function at 
${\cal O}(\alpha_s^3)$ obtained in this work is an analysis of low-$n$ moments
of the $e^+e^-\to c\bar c$ cross section to determine the
$\overline{\mbox{MS}}$ charm quark mass $\overline m_c$ and to investigate the
uncertainty in $\overline m_c$ that arises from the difference of using
fixed-order and contour-improved perturbation theory. For using
contour-improved perturbation theory, which involves integrations of
$\Pi(q^2)$ in the complex $q^2$-plane, it is essential to have the full mass and
$q^2$ dependence of the vacuum polarization function. Such an analysis  was
carried out in Ref.~\cite{Hoang:2004xm} at ${\cal O}(\alpha_s^2)$. Determinations of the charm
quark mass $\overline m_c$ using 
the vacuum polarization function at ${\cal O}(\alpha_s^3)$ in the fixed-order
expansion alone were carried out recently in
Refs.~\cite{Kuhn:2007vp,Boughezal:2006px}. In this paper we discuss 
in detail the reconstruction of the full $q^2$ and mass dependence of the
${\cal O}(\alpha_s^2)$ and ${\cal O}(\alpha_s^3)$ corrections to the 
vacuum polarization function as outlined above. The thorough analysis of
uncertainties in the charm and bottom quark $\overline{\mbox{MS}}$ masses
obtained from low-$n$ moments of the $e^+e^-$ cross section will
be given in a subsequent publication.

The program of this paper is as follows:
In Sec.~\ref{sectionnotation} we set up our notation and in
Sec.~\ref{sectionmethod} we present the basic features of our method for
reconstructing the vacuum polarization function. 
In Sec.~\ref{sectionPilog} we explain details about how logarithmic
contributions in the expansions in the threshold region and for high energies
are incorporated and in Sec.~\ref{sectionPireg} we present how the remaining
non-logarithmic terms are treated. Some of the solutions we obtain have
unphysical properties. Criteria that allow to identify and discard such
solutions are discussed in Sec.~\ref{sectionunphysical}. Numerical analyses
for the ${\cal O}(\alpha_s^2)$ and ${\cal O}(\alpha_s^3)$ contributions of the
vacuum polarization function are given in Secs.~\ref{sectionOas2} and
\ref{sectionOas3}. Our conclusions are given in Sec.~\ref{sectionconclusions}.

\section{Notation}
\label{sectionnotation}

The relation between the normalized $e^+e^-$ cross section $R$ and and the
vacuum polarization function $\Pi$ reads
\begin{align}
\label{eq:spectral-function}
R(q^2) \, = \, 12\pi\,Q_q^2\,\mbox{Im}\,\Pi(q^{2}+i0,m^2)\,,
\end{align}\
where $Q_q$ is the heavy quark electric charge.
The perturbative fixed-order expansion of $\Pi(q^2,m^2)$ has the form
\begin{align}
\label{eq:General-Pi}
\Pi(q^{2},m^2) \, = \,&\, 
\Pi^{(0)}(q^{2},m^2) 
\, + \,\left(\frac{C_F\,\alpha_{s}(\mu^{2})}{\pi}\right)\,
\Pi^{(1)}(q^{2},m^2)
\nn\\[2mm] &
\, + \,
\left(\frac{\alpha_{s}(\mu^{2})}{\pi}\right)^{2}\,
\Pi^{(2)}(q^{2},m^2,\mu^2)
\, + \,
\left(\frac{\alpha_{s}(\mu^{2})}{\pi}\right)^{3}\Pi^{(3)}(q^{2},m^2,\mu^2)
\,+\,\cdots\,,
\end{align}
with the color factor $C_F=4/3$.
We use the on-shell normalization of the vacuum polarization function, where 
\begin{align}
\label{Thomson}
\Pi(0,m^2) \, = \,0\,.
\end{align}
We exclude the so-called singlet contributions where the vacuum polarization
function contains a three-gluon cut. Note that in this work we do not
distinguish between the contributions in $\Pi^{(2)}$ and $\Pi^{(3)}$
proportional to the different SU(3) group theory color factors since there
isn't any compelling technical reason that would make such a distinction
mandatory. This approach neglects the existence of the multi-particle cuts from
diagrams with the insertion of massive fermion loops. Their contribution is
strongly phase-space-suppressed and can be safely ignored for the 
level of accuracy intended in this work.
We emphasize, however, that our approach can be applied to the individual color
contributions as well. 

For the reconstruction of the vacuum
polarization function 
accomplished in this work we use exclusively the pole mass scheme, $m=m_{\rm
  pole}$, since it allows for the most transparent treatment of the
information from the quark pair production threshold. Moreover we use the
choice $\mu=m=m_{\rm pole}$ for the renormalization scale and generally
suppress the $\mu$-dependence of the functions $\Pi^{(i)}$. To simplify the
presentation we frequently use the variable 
\begin{align}
z \, \equiv \, \frac{q^2}{4 m^2}
\,.
\end{align}
For the strong coupling we use $n_f=n_\ell+1$ active running flavors, where
quarks that are heavier than those produced by the current $j^\mu$ are
integrated out and where all $n_\ell$ light flavors are treated as massless. 

The analytic expression for the vacuum polarization functions at 
${\cal O}(\alpha_s)$~\cite{Kallen:1955fb} is an important ingredient of    
our analysis. The corresponding contributions using the notation of
Eq.~(\ref{eq:General-Pi}) have the form 
\begin{align}
\label{eq:Pi01}
\Pi^{(0)} & \, = \, 
\frac{3}{16\pi^{2}}\left[\frac{20}{9}+\frac{4}{3z}-\frac{4(1-z)(1+2z)}{3z}G(z)\right],
\nn \\[2mm] 
\Pi^{(1)} & \, = \, 
\frac{3}{16\pi^{2}}\left[\frac{5}{6}+\frac{13}{6z}-\frac{(1-z)(3+2z)}{z}G(z)+
\frac{(1-z)(1-16z)}{6z}G^{2}(z)\right.
-\,\left.\frac{(1+2z)}{6z}\left(1+2z(1-z)\frac{d}{dz}\right)\frac{I(z)}{z}\right],
\end{align} 
where
\begin{align} 
\label{eq:Gz}
I(z) & \, = \, 
6\Big[\zeta_{3}+4\,\mbox{Li}_{3}(-u)+2\,\mbox{Li}_{3}(u)\Big]-
8\Big[2\,\mbox{Li}_{2}(-u)+\mbox{Li}_{2}(u)\Big]\ln u   
 -2\Big[2\,\ln(1+u)+\ln(1-u)\Big]\ln^{2}u\,, 
\nn \\[2mm]
 G(z) & \, = \, \frac{2\, u\,\ln u}{u^{2}-1}\,,
\quad
\mbox{with}\quad
 u \, \equiv \, \frac{\sqrt{1-1/z}-1}{\sqrt{1-1/z}+1}
\,.
\end{align}
An important application is the determination of the moments $M_n$ in the
fixed-order expansion. Here the pole mass scheme is strongly disfavored since
it contains an ${\cal O}(\Lambda_{\rm QCD})$ renormalon ambiguity that leads to
a quite bad perturbative expansion of the moments. For small values of $n$
this problem can be avoided conveniently by using the
$\overline{\mbox{MS}}$ mass scheme. For the complications that arise for large
values of $n$ see e.g. Refs.~\cite{Hoang:1998uv,Hoang:1999ye}.
In this paper we use the
$\overline{\mbox{MS}}$ running mass $\overline m$ with $n_f=n_\ell+1$ running
flavors for discussions of the moments $M_n$. Using a common renormalization
scale $\mu$ for the mass and the strong coupling, the fixed-order perturbative
expansion of the moments $M_n$ can be written in the form [$l_{m\mu}=\ln(\bar m^2(\mu)/\mu^2)$]
\begin{align}
\label{Mnexpanded}
M_n \, = \, & \,\frac{9}{4}\,\frac{Q_q^2}{(4\bar m^2(\mu))^n}\,
\bigg[ \bar{C}_n^{(0)}
  + \frac{\alpha_s(\mu)}{\pi}
  \left( \bar{C}_n^{(10)} + \bar{C}_n^{(11)}l_{m\mu} \right)
+\left(\frac{\alpha_s(\mu)}{\pi}\right)^2
  \left( \bar{C}_n^{(20)} + \bar{C}_n^{(21)}l_{m\mu}
  + \bar{C}_n^{(22)}l_{m\mu}^2 \right)
\nn \\[2mm] & \hspace{2.5cm}
 +\,\left(\frac{\alpha_s(\mu)}{\pi}\right)^3
  \left( \bar{C}_n^{(30)} + \bar{C}_n^{(31)}l_{m\mu}  + \bar{C}_n^{(32)}l_{m\mu}^2 + 
\bar{C}_n^{(33)}l_{m\mu}^3 \right)\bigg] 
\,,
\end{align}  
adopting the notation of Refs.~\cite{Chetyrkin:1995ii,Chetyrkin:1996cf}.

\section{The Method}
\label{sectionmethod}

The expansions of $\Pi(z)$ in the threshold region $z\simeq 1$ and the
high-energy limit $|z|\to\infty$ involve powers of $\log(1-z)$ and
$\log(-4z)$, respectively. Above production threshold for $z>1$ these logarithmic
terms contribute to the absorptive parts in $\Pi$ that constitute  the
cross section according to Eq.~(\ref{eq:spectral-function}). On the other
hand, the expansion around $z=0$, which is located in 
the Euclidean region, leads to fully analytic terms and admits a usual
Taylor expansion. We want to reconstruct the full $q^2$ and mass dependence of 
$\Pi^{(3)}$ by building functions that incorporate all known
properties of $\Pi^{(3)}$ in the threshold
regime, the high-energy limit and the region around $z=0$. We carry out the
same program also for $\Pi^{(2)}$ using only coefficients in the expansions
that are analogous to the available information for $\Pi^{(3)}$. From the
reconstructed $\Pi^{(3)}$ we can determine previously unknown non-logarithmic
coefficients in the high-energy and the nonrelativistic expansions as well
as the ${\cal O}(\alpha_s^3)$ corrections of the moments $M_n$ for $n\ge 3$. 
Using the reconstructed $\Pi^{(2)}$ function we can test the reliability of these
determinations and find that these coefficients and moments can be determined
remarkably well. 

Following the approach of Ref.~\cite{Chetyrkin:1995ii}, we split
$\Pi^{(2,3)}(z)$ into two parts, 
\begin{align}
\label{Piseparated}
\Pi^{(2,3)}(z) \, = \, \Pi^{(2,3)}_{\rm reg}(z) \, + \, \Pi^{(2,3)}_{\rm log}(z)
\,,
\end{align}
where $\Pi^{(2,3)}_{\rm log}(z)$ are designed such that they contain the
logarithmic terms in the expansions around $z=1$ and for
$|z|\to\infty$. They can be conveniently constructed from the functions
$\Pi^{(1)}$ and $G(z)$ given in Eqs.~(\ref{eq:Pi01}) and (\ref{eq:Gz}) since
the latter readily 
incorporate analytic structures that allow to incorporate the appropriate
threshold and high-energy behavior into $\Pi^{(2,3)}_{\rm log}(z)$. Once
$\Pi^{(2,3)}_{\rm log}(z)$ has been specified, the remaining task is to
construct a Pad\'e approximant for $\Pi^{(2,3)}_{\rm reg}$ that allows to
incorporate the remaining non-logarithmic constraints in the regions $z\simeq
1$, $|z|\to\infty$ and $z\simeq 0$. The general structure of a Pad\'e
approximant $P_{n,m}$ has the form
\begin{align}
\label{Padedef}
P_{n,m}(x) \, = \, 
\frac{\sum_{i=0}^{n}a_{i}x^{i}}{1+\sum_{j=1}^{m}b_{j}x^{j}}
\,,
\end{align}
which means that there are $n+m+1$ coefficients that need to be specified.
Note that the coefficients $a_i$ and $b_j$ are real numbers.
Since $\Pi^{(2,3)}_{\rm reg}$ still has a physical cut for $z>1$ along the
positive real $z$ axis, one cannot use the variable $z$ to formulate the
Pad\'e approximant. A convenient variable to automatically account for this
cut is $\omega$ defined by (see
e.g. Refs.~\cite{Broadhurst:1993mw,Fleischer:1994ef}) 
\begin{align}
\label{eq:omega}
\omega \, = \,
\frac{1-\sqrt{1-z}}{1+\sqrt{1-z}} 
\,,\qquad \qquad
z \, = \, \frac{4\omega}{(1+\omega)^{2}}.
\end{align}
Here, the physical $z$-plane is mapped into the unit-circle of the complex
$\omega$-plane, where approaching the physical cut from the upper (lower)
complex $z$-half-plane corresponds to approaching the upper (lower) semi 
unit-circle in the complex $\omega$-plane. The 
three points $z=(0,1,\pm\infty)$ are conformally mapped onto
$\omega=(0,1,-1)$. Expressed in terms of the variable $\omega$,
$\Pi^{(2,3)}_{\rm reg}$ can therefore be approximated by rational functions
involving the Pad\'e approximant $P_{n,m}(\omega)$. All Pad\'e approximants
that turn out to have unphysical poles inside the unit circle have to be
discarded. In practice some additional restrictive criteria have to be imposed
to avoid an unphysical behavior of $\Pi$ and $R$ due to poles in the Pad\'e
approximant outside the unit circle that are either close to the unit circle
or have a large residue. We discuss these restrictions in
Sec.~\ref{sectionunphysical}.    

It goes without saying that the constructions of $\Pi^{(2,3)}_{\rm log}(z)$ and
the Pad\'e approximant for $\Pi^{(2,3)}_{\rm reg}$ are not unique and that the
resulting reconstructed $\Pi^{(2,3)}$ functions have a dependence on
choices made for their construction. The ambiguity in the
procedure therefore needs to be quantified by accounting for variations in
the construction. While in Ref.~\cite{Chetyrkin:1995ii,Chetyrkin:1996cf}
variations coming from different 
choices for $P_{n,m}$ were included for the error estimate, we include in our
work in addition continuous variations in the construction of $\Pi^{(2,3)}_{\rm log}$.
We test the reliability of the method by determining properties of
$\Pi^{(2)}$ that are precisely known analytically, but that have not been
incorporated for the construction of the approximation for $\Pi^{(2)}$.

\section{Designing $\Pi_{\rm log}$}
\label{sectionPilog}

To determine $\Pi_{\rm log}^{(2,3)}$ we need to account for the
logarithmic terms that arise in $\Pi^{(2,3)}$ in the threshold region
$z\to 1$ and in the high-energy limit $|z|\to\infty$. To facilitate
the presentation it is convenient to write
\begin{align}
\label{Pilogdef}
\Pi_{\rm log}^{(2,3)}(z) \, = \,
\Pi_{\rm thr}^{(2,3)}(z) \, + \,
\Pi_{\rm inf}^{(2,3)}(z) \, + \,
\Pi_{\rm zero}^{(2,3)}(z)
\,,
\end{align}
where $\Pi_{\rm thr}^{(2,3)}$ and $\Pi_{\rm inf}^{(2,3)}$ are designed
to account for the logarithmic terms at threshold and at high
energies, respectively, and $\Pi_{\rm zero}^{(2,3)}$ incorporates
subtractions that ensure a physical behavior at $z=0$. 
\\[5mm]

\newpage
\noindent
{\it Threshold Logarithms.} We start by presenting the expansions of
$\Pi^{(1,2,3)}(z)$ and $G(z)$ in the threshold limit $z\to 1$
keeping terms up to NNLO in the expansion in $\sqrt{1-z}$\,:
\begin{align}
\label{Pithresh}
\Pi^{(1)}(z) \, = \,& \,-0.1875\ln(1-z)-0.314871 +0.477465\sqrt{1-z}
\nn\\[1mm] &
+\Big(0.354325 + 0.125 \ln(1-z)\Big)(1-z) 
+ {\cal O}\Big((1-z)^{3/2}\Big)
\,,
\nn\\[1mm]
\Pi^{(2)}(z) \, = \,& 
\frac{1.72257}{\sqrt{1-z}}
+(0.34375-0.0208333 n_{\ell})\ln^{2}(1-z)+(0.0116822 n_{\ell}
+1.64058 )\ln(1-z)
+ K^{(2)}
\nn\\[1mm] & 
+\Big(\!
-0.721213 - 0.0972614 n_\ell +  3.05433 \ln(1-z)\Big)\sqrt{1-z} 
\, + \, {\cal O}\Big((1-z)\Big)
\,,
\nn\\[1mm]
\Pi^{(3)}(z) \, = \,&
\frac{2.63641}{1-z}+\frac{0.678207 n_{\ell}-27.2677}{\sqrt{1-z}}
+(0.57419 n_{\ell}-9.47414)\,\frac{\log(1-z)}{\sqrt{1-z}}
\nn\\[1mm] &
  +(-0.00231481 n_{\ell}^{2}+0.0763889 n_{\ell}-0.630208)\log^{3}(1-z)
\nn\\[1mm] &
+(0.00194703 n_{\ell}^{2}+0.0312341 n_{\ell}+1.3171)\log^{2}(1-z)
\nn\\[1mm] &
+(-0.0690848 n_{\ell}^{2}+2.37068 n_{\ell}-17.6668)\log(1-z)
\, + \, K^{(3)}
+ {\cal O}\Big((1-z)^{1/2}\Big)
\,,
\nn\\[1mm]
G(z) \, = \,& \frac{\pi}{2\sqrt{1-z}} -  1  
+ \frac{\pi\sqrt{1-z}}{4}
\, + \, {\cal O}\Big((1-z)\Big)
\,.
\end{align} 
To avoid cluttering we show the various coefficients for $\Pi^{(1,2,3)}(z)$
in numerical 
form, but keep the number $n_\ell$ of light quark flavors as a
variable. The expansions of $\Pi^{(1)}$ and $G$ are known from their
exact expressions given in Eqs.~(\ref{eq:Pi01}) and (\ref{eq:Gz})
while the expansion for $\Pi^{(2)}$ can be derived from the results
for $R$ in the threshold region computed in Ref.~\cite{Czarnecki:1997vz}. The  
expansion for $\Pi^{(3)}$ is obtained from the NNLO threshold cross
section factorization formula for $R$ within NRQCD first derived in
Ref.~\cite{Hoang:1997sj,Hoang:1998xf} (see also Ref.~\cite{Hoang:2001mm}). The
result was 
later confirmed by many other groups~\cite{Hoang:2000yr}. Note that within
NRQCD it is the standard convention that 
only the $n_\ell$ light quark species contribute to the running of the
strong coupling. Switching to $n_f=n_\ell+1$ running flavors affects
the coefficient of the term $\propto\ln(1-z)$ in
$\Pi^{(3)}$. All other coefficients shown in Eq.~(\ref{Pithresh}) are
unaffected. We also 
note that the singlet contributions to the vacuum polarization
function only affect the threshold expansion at N${}^4$LO in the
expansion in $\sqrt{1-z}$ and do not contribute at the order we
consider here.\footnote{
Within NRQCD the dominant effect of the singlet contributions is
associated to 4-quark operators with a Wilson coefficient that
incorporates the hard effects of the 3-gluon annihilation. This
operator leads to a momentum space potential $\propto
\alpha_s^3/m^2$.} 
The constant terms $K^{(2,3)}$ that appear in the nonrelativistic
expansion of  $\Pi^{(2,3)}$ have not yet been computed from Feynman
diagrams. As we show in Secs.~\ref{sectionOas2} and \ref{sectionOas3} 
they can be determined from the reconstructed vacuum polarization
function based on the method described in Sec.~\ref{sectionmethod}. 

\begin{table}[h]
\begin{center}
\renewcommand{\arraystretch}{1.5}
\begin{tabular}{|c|c|}
\hline
$\ln^{m}(1-z)$ & $[\Pi^{(1)}(z)]^m$
\tabularnewline \hline 
\mbox{\hspace{0.3cm}} $(1-z)^{-n/2}\ln^{m}(1-z)$ 
\mbox{\hspace{0.3cm}} & $[G(z)]^n [\Pi^{(1)}(z)]^m$ 
\tabularnewline \hline 
$(1-z)^{n/2}\ln^{m}(1-z)$ & 
\mbox{\hspace{0.3cm}}
$(1-z)^n [G(z)]^n [\Pi^{(1)}(z)]^m$
\mbox{\hspace{0.3cm}}
\tabularnewline \hline 
\end{tabular}
\caption{First column: Logarithmic terms that arise in the expansion of
  $\Pi^{(2,3)}(z)$ close to threshold where $z\approx 1$. Second column:
  Corresponding functions used in the construction of $\Pi^{(2,3)}_{\rm
    thr}(z)$. 
\label{tab:thresh}}
\end{center}
\end{table}

To construct $\Pi_{\rm thr}^{(2,3)}$ we have to find appropriate
functions that account for the different combinations of the logarithmic
term $\ln(1-z)$  and powers of $\sqrt{1-z}$ that appear in
Eqs.~(\ref{Pithresh}). A convenient
choice is given in Tab.~\ref{tab:thresh}, and leads to 
\begin{align}
\label{Pithreshansatz}
\Pi_{\mathrm{thr}}^{(2)}(z) 
\,=\, &\,
A_{0}^{(2)}\frac{1+a_0^{(2)}\, z}{z}\left[\Pi^{(1)}(z)\right]^{2}
+ A_{1}^{(2)}\Pi^{(1)}(z)
+A_{2}^{(2)}(1-z)G(z)\Pi^{(1)}(z)
\,,
\nn\\[2mm]
\Pi_{\mathrm{thr}}^{(3)}(z)
\, =\, &\,
A_{0}^{(3)}\dfrac{1+a_0^{(3)}\, z}{z}\left[\Pi^{(1)}(z)\right]^{3}   
+A_{1}^{(3)}\left[\Pi^{(1)}(z)\right]^{2}+A_{2}^{(3)} \dfrac{1+a_2^{(3)}\, z}{z}G(z)\Pi^{(1)}(z)
+A_{3}^{(3)}\Pi^{(1)}(z)
\,.
\end{align}
The coefficients $A_i^{(2,3)}$ can be unambiguously determined from
the expressions shown in Eqs.~(\ref{Pithresh}). Obviously the choices in
Tab.~\ref{tab:thresh} are not 
unique. To have some quantitative way to account for this source of
uncertainty we have multiplied the functions related to the highest power of
$\ln(1-z)$ in different orders in the expansion in $\sqrt{1-z}$ by a 
term $(1-a^{(2,3)}_i\,z)/z$, where the $a^{(2,3)}_i$ are free
parameters. Since the construction becomes singular for $a^{(2,3)}_i=-1$, we
exclude this value and use variations  in the ranges $a^{(2,3)}_i \ge0$ and
$a^{(2,3)}_i \le -2$. 
Note that for $|a^{(2,3)}_i|\to\infty$ the term
$(1-a^{(2,3)}_i\,z)/z$ becomes $z$-independent and the results for 
$\Pi^{(2,3)}_{\rm thr}$ become
independent of the $a^{(2,3)}_i$.
We note that given the functions in Tab.~\ref{tab:thresh} it is
straightforward to account for even higher terms in the expansion in the
threshold limit for the construction of $\Pi_{\rm thr}^{(2,3)}$.
\\[5mm]

\noindent
{\it High-Energy Logarithms.} The expansions of $\Pi^{(1,2,3)}(z)$ and $G(z)$
in the high-energy limit $|z|\to\infty$ read
\begin{align}
\label{Pihigh}
\Pi^{(1)}(z) \, = \,& 
-\,0.018998\log(-4z)-0.075514- 0.056993 \frac{\ln(-4z)}{z}
\nn\\[1mm] &
+ \,\frac{0.023628}{z^{2}}- 0.014248 \frac{\ln^{2}(-4z)}{z^{2}}
-  0.011874 \frac{\ln(-4z)}{z^{2}}
+\,{\cal O}(z^{-3})
\,,
\nn\\[1mm]
\Pi^{(2)}(z) \, = \,&  \,
(0.034829 - 0.0021109 n_f)\ln^{2}(-4z)
+(-0.050299 + 0.0029205 n_f)\ln(-4z)
+ H^{(2)}_0
\nn\\[1mm] &
+ (0.18048 - 0.0063326 n_f)\frac{\ln^{2}(-4z)}{z}
+ (-0.59843 + 0.027441 n_f)\frac{\ln(-4z)}{z}
+ \frac{H^{(2)}_1}{z}
\nn\\[1mm] &
+ (0.042745 - 0.0010554 n_f)\frac{\ln^3(-4z)}{z^2} 
+ (-0.10132 + 0.0058049 n_f)\frac{\ln^2(-4z)}{z^2} 
\nn\\[1mm] &
+ (-0.48134 + 0.032065 n_f)\frac{\ln(-4z)}{z^2} 
+ \frac{H^{(2)}_2}{z^2}
+\,{\cal O}(z^{-3})
\,,
\nn\\[1mm]
\Pi^{(3)}(z) \, = \,&   \,
 (-0.063853 + 0.0077398 n_f-0.00023454\ n_f^2) \ln^{3}(-4z)
\nn\\[1mm] &
+ (0.21906 - 0.026441 n_f + 0.0004867 n_f^2) \ln^{2}(-4z)
\nn\\[1mm] &
+ (-0.46209 + 0.10679n_f - 0.0021837n_f^2)\ln(-4z)
+ H^{(3)}_0
\nn\\[1mm] &
+ (-0.45120+ 0.035885 n_f - 0.00070362 n_f^2)\frac{\ln^{3}(-4z)}{z}
+ (3.0848 - 0.26016 n_f +  0.0045735 n_f^2) \frac{\ln^{2}(-4z)}{z}
\nn\\[1mm] &
+ (-6.6516 + 0.78237 n_f - 0.0146587 n_f^2) \frac{\ln(-4z)}{z}
+ \frac{H^{(3)}_1}{z}
\nn\\[1mm] &
+ (-0.10152 + 0.0060687 n_f - 0.000087952 n_f^2) \frac{\ln^{4}(-4z)}{z^{2}}
\nn\\[1mm] &
+ (0.57013 - 0.044856 n_f +  0.0006743 n_f^2) \frac{\ln^{3}(-4z)}{z^{2}}
+ (0.17822 + 0.038525 n_f -  0.00088394 n_f^2) \frac{\ln^{2}(-4z)}{z^{2}}
\nn\\[1mm] &
+ (-8.8712 + 1.0393 n_f - 0.026019 n_f^2) \frac{\ln(-4z)}{z^{2}}
+ \frac{H^{(3)}_2}{z^2}
+\,{\cal O}(z^{-3})
\,,
\nn\\[1mm]
G(z) \, = \,& 
-\frac{\log(-4z)}{2z}
+\frac{1-\log(-4z)}{4z^{2}}
+\,{\cal O}(z^{-3})
\,.
\end{align} 
The expansions for $\Pi^{(1)}$ and $G$ are known from the exact expressions
given in Eqs.~(\ref{eq:Pi01}) and (\ref{eq:Gz}), while the expansion for
$\Pi^{(2)}$ was taken from Ref.~\cite{Chetyrkin:1994ex}. Note that many
orders in high-energy expansion are known for
$\Pi^{(2)}$~\cite{Chetyrkin:1997qi}, but we only consider in this 
work terms up to order $1/z^2$, since our analysis for $\Pi^{(2)}$ mainly serves
as a testing ground for the application to $\Pi^{(3)}$. The expansion for
$\Pi^{(3)}$ was obtained in Refs.~\cite{Chetyrkin:2000zk}. At ${\cal
O}(\alpha_s^2)$ the non-logarithmic coefficients $H^{(2)}_{0,1}$ are known
analytically~\cite{Chetyrkin:1996cf} and read
\begin{align}
\label{HOas}
H^{(2)}_0 \, = \, & -0.73628 + 0.037645 n_f
\,,
\nn\\[2mm]
H^{(2)}_1 \, = \, & -0.30324 + 0.029002 n_f
\,.
\end{align}
At ${\cal O}(\alpha_s^3)$ the non-logarithmic coefficients $H^{(3)}_{0,1}$ 
have not been computed from Feynman diagrams in the literature before.  As we
show in Secs.~\ref{sectionOas2} and \ref{sectionOas3} 
they can be determined from the reconstructed vacuum polarization
function $\Pi^{(3)}$.

\begin{table}[ht]
\begin{center}
\renewcommand{\arraystretch}{1.5}
\begin{tabular}{|c|c|}
\hline
$\ln^{n}(-4z)$ & $(1-z)^n\,[G(z)]^n$
\tabularnewline \hline 
\mbox{\hspace{0.3cm}}
$\frac{1}{z}\ln^{n}(-4z)\,,(n>1)$ 
\mbox{\hspace{0.3cm}}
& 
\mbox{\hspace{0.3cm}}
$(1-z)^{n-1}\,[G(z)]^n$ 
\mbox{\hspace{0.3cm}}
\tabularnewline \hline 
$\frac{1}{z}\ln(-4z)$ &  $\frac{1-z}{z}\,G(z)$
\tabularnewline \hline 
$\frac{1}{z^2}\ln(-4z)$ & $\frac{1-z}{z^2}\,G(z)$
\tabularnewline \hline 
$\frac{1}{z^2}\ln^2(-4z)$ & $\frac{1-z}{z}\,[G(z)]^2$
\tabularnewline \hline 
$\frac{1}{z^2}\ln^3(-4z)$ & $\frac{(1-z)^2}{z}\,[G(z)]^3$
\tabularnewline \hline 
$\frac{1}{z^2}\ln^4(-4z)$ & $(1-z)^2\,[G(z)]^4$
\tabularnewline \hline 
\end{tabular}
\caption{First column: Logarithmic terms that arise in the high-energy
  expansion of $\Pi^{(2,3)}(z)$ where $|z|\to\infty$. Second column:
  Corresponding functions used in the construction of $\Pi^{(2,3)}_{\rm inf}(z)$.
\label{tab:high}}
\end{center}
\end{table}

To construct $\Pi^{(2,3)}_{\rm inf}(z)$
we have to find functions that can account for the different combinations of
powers of $\ln(-4z)$ and of powers of $1/z$ that arise in the expansions of
Eq.~(\ref{Pihigh}). A convenient
choice is given in Tab.~\ref{tab:high}. Our guideline for including the
factors of $(1-z)^i$ is to ensure that the functions are constant or
$\sim\sqrt{1-z}$ in the threshold limit $z\to 1$. This leads to 
\begin{align}
\label{Pihighansatz}
\Pi_{\rm{inf}}^{(2)} 
\, = \,& 
B_{0}^{(2)}\frac{1+b_{0}^{(2)}z}{z}(1-z)^{2}G(z)^{2}+\left(
  B_{10}^{(2)}+\frac{B_{11}^{(2)}}{z}\right)
\frac{1+b_{1}^{(2)}z}{z}(1-z)G(z)^{2}
\nn  \\[2mm] &  
+ \left(B_{30}^{(2)}+\frac{B_{31}^{(2)}}{z}+\frac{B_{32}^{(2)}}{z^2}\right)(1-z)G(z)
+ B_{4}^{(2)}\frac{(1-z)^{2}}{z}G(z)^{3}
,
\nn \\[2mm]
\Pi_{\rm{inf}}^{(3)} 
\, = \, &
\frac{1+b_{0}^{(3)}z}{z}\left(B_{00}^{(3)}
+\frac{B_{01}^{(3)}}{z}\right)(1-z)^{3}G(z)^{3}+B_{1}^{(3)}(1-z)^{2}G^{4}(z)
+\frac{1+b_{2}^{(3)}z}{z}\left(B_{20}^{(3)}
+\frac{B_{21}^{(3)}}{z}\right)(1-z)^{2}G(z)^{3}
\nn \\[2mm] & 
 +B_{30}^{(3)}(1-z)^{2}G(z)^{2}+\left(B_{40}^{(3)}+\frac{B_{41}^{(3)}}{z}\right)(1-z)G(z)^{2}
+\left(B_{50}^{(3)}+\frac{B_{51}^{(3)}}{z}+\frac{B_{52}^{(3)}}{z^{2}}\right)(1-z)G(z)
\,,
\end{align}
where the coefficients $B^{(n)}_i$ can be determined unambiguously from the
conditions in Eqs.~(\ref{Pithresh}) and (\ref{Pihigh}). In analogy to
$\Pi_{\rm thr}^{(2,3)}$ we 
have included modification factors $(1+b^{(2,3)}_i\,z)/z$ for the functions that
are related to highest-power logarithmic terms at each order in the $1/z$
expansion. In $\Pi^{(3)}_{\rm high}$ we have a common modification factor for
the functions 
related to the terms $\ln^3(-4 z)/z$ and   $\ln^3(-4 z)/z^2$. For the
parameters $b^{(2,3)}_i$ the choice $b^{(2,3)}_i=0$ is excluded because in this
case the construction becomes singular. For our analysis we adopt variations
in the ranges $|b^{(2,3)}_i|\ge 1$. 
Using functions along the lines of Tab.~\ref{tab:thresh} it is
straightforward to account for even higher terms in the high-energy expansion
in the  for the construction of $\Pi_{\rm inf}^{(2,3)}$.

Note that for the determination of the coefficient $A_i^{(2,3)}$ and $B_i^{(2,3)}$
one first fixes the constants $a^{(2,3)}_i$ and $b^{(2,3)}_i$ in the modification
functions and then solves a set of linear equations. 
\\[5mm]

\noindent
{\it Subtractions at $q^2=0$.} 
There are singularities $\sim 1/z$ and $\sim 1/z^2$ in 
$\Pi_{\rm thr}^{(2,3)}(z)$ and $\Pi_{\rm inf}^{(2,3)}(z)$ that arise 
in the limit $z\to 0$. They are a consequence of the functions used to
construct  $\Pi_{\rm thr}^{(2,3)}(z)$ and $\Pi_{\rm inf}^{(2,3)}(z)$.
These singularities lead to unphysical behavior and need to be subtracted. For
this task we define the function
\begin{align}
\label{Pizero}
\Pi_{\rm zero}^{(2,3)}(z) \, = \,
S_0^{(2,3)} \, + \, \frac{S_1^{(2,3)}}{z} 
\, + \,\frac{S_2^{(2,3)}}{z^2}
\,.
\end{align}
After the coefficients $A_i^{(2,3)}$ and $B_i^{(2,3)}$ have been computed, the
coefficients $S_{0,1,2}^{(2,3)}$ are determined such that
\begin{align}
\label{Thomson2}
\Pi^{(2,3)}_{\rm log}(0) \, = \, 0
\,.
\end{align}
Note that it is not mandatory to fix $S_0^{(2,3)}$ in this way, and that our
approach is independent of the choice for $S_0^{(2,3)}$. However, to satisfy
Eq.~(\ref{Thomson}) it is  
convenient for the purpose of presentation to impose the
condition~(\ref{Thomson2}) and also the relation $\Pi^{(2,3)}_{\rm reg}(0)=0$.

\section{Designing $\Pi_{\rm reg}$}
\label{sectionPireg}

The terms $\Pi_{\rm reg}^{(2,3)}$ in Eq.~(\ref{Piseparated}) have to account
for the non-logarithmic conditions in the expansion at the threshold and at
high energies, and  for the coefficients that arise in the expansion around
$z=0$. We start by presenting the small-$z$ expansion of $\Pi^{(2)}$ and
$\Pi^{(3)}$:
\begin{align}
\label{Pismallz}
\Pi^{(2)} \, = \,& 
(0.719976- 0.0296233 n_\ell) z + (0.698894- 
    0.0275334 n_\ell) z^2 + (0.637986- 0.0240088 n_\ell) z^3 
\nn \\[1mm] &
+ (0.584109- 
    0.0211621 n_\ell) z^4 + (0.539450- 0.0189263 n_\ell) z^5 
 + (0.502392- 0.0171420 n_\ell) z^6 
\nn\\[1mm] &+ (0.471258- 0.0156884 n_\ell) z^7
+\,{\cal O}(z^8)
\,,
\nn\\[2mm]
\Pi^{(3)} \, = \,& 
(10.6103- 1.30278 n_\ell + 0.0282783 n_\ell^2)z
+(10.4187- 1.12407 n_\ell + 0.0223706 n_\ell^2)z^2
\,+\,{\cal O}(z^3)
\,.
\end{align}
The coefficients for  $\Pi^{(2)}$ were computed in
Refs.~\cite{Chetyrkin:1995ii,Chetyrkin:1996cf}. Recently 
the coefficients for  $\Pi^{(2)}$ have even been determined up to order
$z^{30}$~\cite{Maier:2007yn,Boughezal:2006uu}. For $\Pi^{(3)}$ the
coefficient of order $z$ was computed in
Refs.~\cite{Chetyrkin:2006xg,Boughezal:2006px}, and the
coefficient of order $z^2$ was given in Ref.~\cite{Maier:2008he}. The
coefficients of order $z^n$ with $n\ge 
3$ have not yet been computed from Feynman diagrams. However, they can be
determined from the reconstructed function $\Pi^{(3)}$ as we show in
Secs.~\ref{sectionOas2} and \ref{sectionOas3}. 
\\[5mm]

\noindent
{\it Designing $\Pi_{\rm reg}^{(2)}$.} 
We start exemplarily with the construction of $\Pi_{\rm reg}^{(2)}$. Close to
threshold $\Pi^{(2)}$ exhibits the Coulomb singularity $\sim 1/\sqrt{1-z}$,
see Eq.~(\ref{Pithresh}). To avoid that the Pad\'e approximant contains
explicitly this singularity, we use two different methods:
\begin{itemize}
\item[(i)]
We relate the Pad\'e approximant $P(\omega)$ to $f(z)\Pi_{\rm reg}^{(2)}$,
where $f(z\approx 1)\sim \sqrt{1-z}$. The coefficient of the Coulomb
singularity is implemented through a condition on $P(1)$.
\item[(ii)] 
We use the relation
\begin{align}
\frac{\pi^{2}}{9}\, G(z\approx 1) 
\, =\, 
\frac{\pi^{3}}{18\sqrt{1-z}} \, + \, \ldots
\end{align}
and account for the Coulomb singularity by adding the function
$\frac{\pi^{2}}{9}G(z)$ to 
$\Pi^{(2)}_{\rm log}$. The Pad\'e approximant $P$ is not affected by the
Coulomb singularity.
\end{itemize}
The numerical differences that result from
these two methods of implementing the Coulomb singularity constitute
another tool for quantifying the uncertainties inherent to our approach. 

For method (i) the expression we use for the relation between the Pad\'e
approximant $P(\omega)$ and $\Pi^{(2)}_{\rm reg}$ reads 
\begin{align}
\label{Pa}
P(\omega) \, = \,
\frac{1-\omega}{(1+\omega)^{2}}
\left[\Pi_{\mathrm{reg}}^{(2)}(z)-\Pi_{\mathrm{reg}}^{(2)}(-\infty)\right]
\,,
\end{align}
where $\frac{1-\omega}{(1+\omega)^2}\sim \sqrt{1-z}$\, for\, $z\to 1$. 
A similar relation was also used in Ref.~\cite{Chetyrkin:1996cf}.
Since the
prefactor grows linearly with $z$, $P(-1)$ is a finite number. Some comments
are in order concerning the term $\Pi_{\rm reg}(-\infty)$ that appears in
Eq.~(\ref{Pa}) and also in the analogous relations~(\ref{Pb}) and (\ref{Pc})
that follow below. From the conditions 
$\Pi(0)=\Pi_{\rm log}(0)=\Pi_{\rm reg}(0)=0$ it is easy to see that 
\begin{align}
\label{PzeroPi}
P(0) \, = \, -\,\Pi_{\rm reg}(-\infty)
\,.
\end{align}
Thus in case that $\Pi_{\rm reg}(-\infty)$ is known and taken as an input,
Eq.~(\ref{PzeroPi}) represents a condition that is imposed on the Pad\'e
approximant $P$. On the other hand, if $\Pi_{\rm reg}(-\infty)$ is unknown or
not taken as an input, it can be determined from Eq.~(\ref{PzeroPi}) once the
Pad\'e approximant has been fixed from other conditions. We show in
Secs.~\ref{sectionOas2} and \ref{sectionOas3} that this allows to determine
the high energy constants $H^{(2,3)}_0$ with small 
uncertainties. From Eqs.~(\ref{Pa}), (\ref{PzeroPi}) and (\ref{Piseparated})
the vacuum polarization function $\Pi^{(2)}$ is recovered from the relation
\begin{align}
\label{Pia}
\Pi^{(2)}(z) \, = \,
\frac{(1+\omega)^{2}}{1-\omega}\,P(\omega) \,- \, P(0)
\, + \, \Pi_{\mathrm{log}}^{(2)}(z)\,.  
\end{align}
From Eqs.~(\ref{Pa}) it is now straightforward to determine the conditions on
the Pad\'e approximant $P(\omega)$ from the non-logarithmic constraints on
$\Pi(z)$ in the threshold and the  high-energy regions and from the
coefficients in the expansion around $z=0$. Additional constraints on $P$
arise from the fact 
that in the limit $|z|\to\infty$ the first term on the RHS of Eq.~(\ref{Pia})
can exhibit odd power terms $\sim 1/z^{(2n+1)/2}$ with $n=1,2,\ldots$, which do
not exist in the high-energy expansion of the vacuum polarization function.
It is reasonable to exclude such terms up to order $1/z^{(2n+1)/2}$ when the
information from the high-energy expansion up to order $1/z^n$ is accounted
for. For example, excluding terms $\sim 1/z^{3/2}$ in Eq.~(\ref{Pia}) leads to
the constraint $P(-1)-2P^\prime(-1)=0$, where $P^\prime$ refers to the
derivative of $P(\omega)$ with respect to $\omega$. Excluding also terms $\sim
1/z^{5/2}$ leads to the condition 
$3P(-1)-9P^{\prime\prime}(-1)+2P^{\prime\prime\prime}(-1)=0$.
The various conditions on $P$ lead to a complicated non-linear set of
equations for the coefficients of the Pad\'e approximant in
Eqs.~(\ref{Padedef}), which we do not present explicitly here. These equations
frequently have multiple solutions and are most conveniently tackled numerically. 

For method (ii), where the Coulomb singularity is treated in 
$\Pi_{\rm log}^{(2)}$ the relation between the Pad\'e approximant $P(\omega)$
and $\Pi_{\rm reg}^{(2)}(z)$ reads
\begin{align}
\label{Pb}
P(\omega) \, = \,
\frac{1}{(1+\omega)^{2}}
\left[\Pi_{\mathrm{reg}}^{(2)}(z)-\Pi_{\mathrm{reg}}^{(2)}(-\infty)\right]
\,.
\end{align} 
A similar relation was also used in Ref.~\cite{Chetyrkin:1996cf}.
Here, excluding terms of order $1/z^{3/2}$ and $1/z^{5/2}$ for $|z|\to\infty$
corresponds to the conditions $P(-1)-P^\prime(-1)=0$ and 
$6P(-1)-6P^{\prime\prime}(-1)+P^{\prime\prime\prime}(-1)=0$,
respectively. The vacuum polarization function is then recovered from the
relation 
\begin{align}
\label{Pib}
\Pi^{(2)}(z) \, = \, 
(1+\omega)^{2}\,P(\omega) \, - \, P(0)
\, + \, \Pi_{\mathrm{log}}^{(2)}(z)\,.
\end{align}
\\[0mm]

\noindent
{\it Designing $\Pi_{\rm reg}^{(3)}$.}
The construction of $\Pi_{\rm reg}^{(3)}$ proceeds in a similar way. At ${\cal
  O}(\alpha_s^3)$ the vacuum polarization function has a Coulomb singularity
$\sim 1/(1-z)$. For method (i) this singularity is incorporated in 
$\Pi_{\rm reg}^{(3)}$ and the relation between $P(\omega)$ and 
$\Pi_{\rm reg}^{(3)}$ reads
\begin{align}
\label{Pc}
P(\omega) \, = \, 
\left(\frac{1-\omega}{1+\omega}\right)^{2}
\left[\Pi_{\mathrm{reg}}^{(3)}(z)-\Pi_{\mathrm{reg}}^{(3)}(-\infty)\right]
\,,
\end{align}
where $(1-\omega)^2\sim(1-z)$ for $z\to 1$. 
Excluding terms of order $1/z^{3/2}$ and $1/z^{5/2}$ for $|z|\to\infty$
corresponds to the conditions $P^\prime(-1)=0$ and
$3P^{\prime\prime}(-1)-P^{\prime\prime\prime}(-1)=0$, 
respectively. The vacuum polarization function is recovered from the relation
\begin{align}
\label{Pic}
\Pi^{(3)}(z) \, = \, 
\left(\frac{1+\omega}{1-\omega}\right)^{2}\,P(\omega)
\, - \, P(0) \, + \, \Pi_{\mathrm{log}}^{(3)}(z)\,.
\end{align} 

For method (ii) we add the function $8\zeta_3[G(z)]^2/9$ to 
$\Pi^{(3)}_{\rm log}$. Since $8\zeta_3[G(z)]^2/9\to 2\,\pi^2\zeta_3/[9(1-z)]$
for $z\to 1$ the 
Coulomb singularity is therefore accounted for in $\Pi^{(3)}_{\rm log}$. The 
relation between the Pad\'e approximants and $\Pi^{(3)}_{\rm reg}$ has then
the same form as Eq.~(\ref{Pa}) with $\Pi^{(2)}_{\rm reg}$ replaced by
$\Pi^{(3)}_{\rm reg}$. The relation for the $\Pi^{(3)}(z)$ has the same form
as Eq.~(\ref{Pia}) with $\Pi^{(2)}_{\rm log}$ replaced by 
$\Pi^{(3)}_{\rm log}$. The relations imposed on $P$ to exclude terms of order
$1/z^{3/2}$ and $1/z^{5/2}$ are then also $P(-1)-2P^\prime(-1)=0$ and 
$3P(-1)-9P^{\prime\prime}(-1)+2P^{\prime\prime\prime}(-1)=0$,
respectively.

\section{Discarding Unphysical Solutions}
\label{sectionunphysical}

For the reconstructed functions $\Pi^{(2)}$ and $\Pi^{(3)}$ we have several
types of variations that can be implemented into the construction and which we
can use to quantify numerically the uncertainty of the results. Apart from
the two ways to account for the Coulomb singularity described as methods (i)
and (ii) in the previous section, we have also implemented modification
factors in  Eqs.~(\ref{Pithreshansatz})   and (\ref{Pihighansatz}) that allow
us to scan  
over a continuous set of functions within $\Pi^{(2,3)}_{\rm log}$. Once the
modification functions are fixed there are in general several possible choices 
one can use for the Pad\'e approximants $P_{m,n}$ with $n+m$ being fixed by
the number of conditions one imposes on $\Pi^{(2,3)}_{\rm reg}$. 
The resulting solutions for the  Pad\'e approximants can, however, have
properties that lead to an unphysical and pathological behavior for $\Pi$ and
$R$. Such solutions need to be discarded for a meaningful phenomenological
analysis~\cite{Chetyrkin:1996cf}.   

An obvious restriction concerns solutions for $P_{m,n}(\omega)$ that lead to
poles in $\Pi^{(2,3)}$ in the complex $\omega$-plane inside the unit
circle.\footnote{For Pad\'e approximants of the form $P_{k,0}$ (Taylor-like) 
such poles do not exist and none of the solutions is
discarded.}  
These solutions are unacceptable and we discard them right away 
because such poles lead to an unphysical analytic structure. A more subtle
situation arises for solutions 
with poles in  $\Pi^{(2,3)}$ in the upper complex $\omega$-half-plane that are
outside the unit circle, but are either 
close to the unit circle or have a large residue. Although the analytic
structure of such solutions is not a priori wrong, we still discard
such solutions if they lead to an unphysical resonance-like structure in the
cross section $R$. To have a quantitative criterion that can be implemented
easily automatically we compute for every pole in $\Pi_{\rm reg}$ in the upper
complex $\omega$-half-plane the so called {\it pole factor}
\begin{align}
\label{polefactor}
\rho \, = \, \frac{|\mbox{Res}_{\Pi}(\omega_{\rm pole})|}{|\omega_{\rm pole}|-1}
\,,
\end{align}
where $\omega_{\rm pole}$ is the location of the pole in the complex
$\omega$-plane and $\mbox{Res}_{\Pi}(\omega_{\rm pole})$ the residue of
$\Pi^{(2,3)}$ at $\omega_{\rm pole}$. If $|\omega_{\rm pole}|$ is close to
unity or if the residue is large, the pole factor becomes big and a
resonance-like structure can arise in $R$. We discard solutions when 
$\rho > \rho_{0}$. 
For our analysis we found that the choice $\rho_0=2.8$ represents a reasonable
restriction for $\Pi^{(2)}$, while for $\Pi^{(3)}$ we use $\rho_0=30$. For the
vacuum polarization function $\Pi^{(3)}$ a larger value for $\rho_0$ is
used since such poles arise predominantly in the threshold region close to
$\omega=1$. Here $\mbox{Im}[\Pi^{(3)}]$ is substantially larger than
$\mbox{Im}[\Pi^{(2)}]$ due to the bigger size of its Coulomb singularity,
see Eq.~(\ref{Pithresh}). Given the set of solutions for $\Pi^{(2)}$ that pass
the restrictions described above we can analyze how well these
solutions reproduce other well-known properties of $\Pi^{(2)}$.

\section{Analysis for the Vacuum Polarization at ${\cal O}(\alpha_s^2)$}
\label{sectionOas2}

The purpose of this section is two-fold. First we demonstrate the
reliability of our approach for its application to $\Pi^{(3)}$ by testing it
with the rather well-known  
${\cal O}(\alpha_s^2)$ vacuum polarization function $\Pi^{(2)}$ and, second,
we determine the previously unknown constant $K^{(2)}$ that appears in the
nonrelativistic expansion of $\Pi^{(2)}$ close to the threshold, see
Eq.~(\ref{Pithresh}). 

To demonstrate the reliability of our approach let us reconstruct $\Pi^{(2)}$
using only information from the different expansions that is analogous to the
available information in $\Pi^{(3)}$. Thus we account for the expansions in
the threshold region up to NNLO, in the high-energy region up to order $1/z^2$
and up to order $z^2$ for the expansion around $z=0$. For the construction of
$\Pi^{(2)}_{\rm reg}$ this entails that we account for the first two
coefficients of the expansion around $z=0$, the non-logarithmic term
$\propto\sqrt{1-z}$ in the threshold region and the constraints that terms
$\sim 1/z^{3/2}$ and $\sim 1/z^{5/2}$ are absent for $|z|\to\infty$. We do not
implement the known constants $H^{(2)}_{0,1}$, but we determine them from
the reconstructed $\Pi^{(2)}$. This amounts to 6 constraints on the Pad\'e
approximant for method (i), where the Coulomb singularity is accounted for in
$\Pi^{(2)}_{\rm reg}$, and to 5 constraints on the Pad\'e  approximant for
method (ii), where the Coulomb singularity is accounted for in 
$\Pi^{(2)}_{\rm log}$. Thus we have $n+m=5$ for the Pad\'e approximants
$P_{m,n}$ for method~(i) and $n+m=4$ for the Pad\'e approximants for 
method~(ii).   

\begin{figure}[ht]
\begin{center}
\epsfxsize=\textwidth
\epsffile{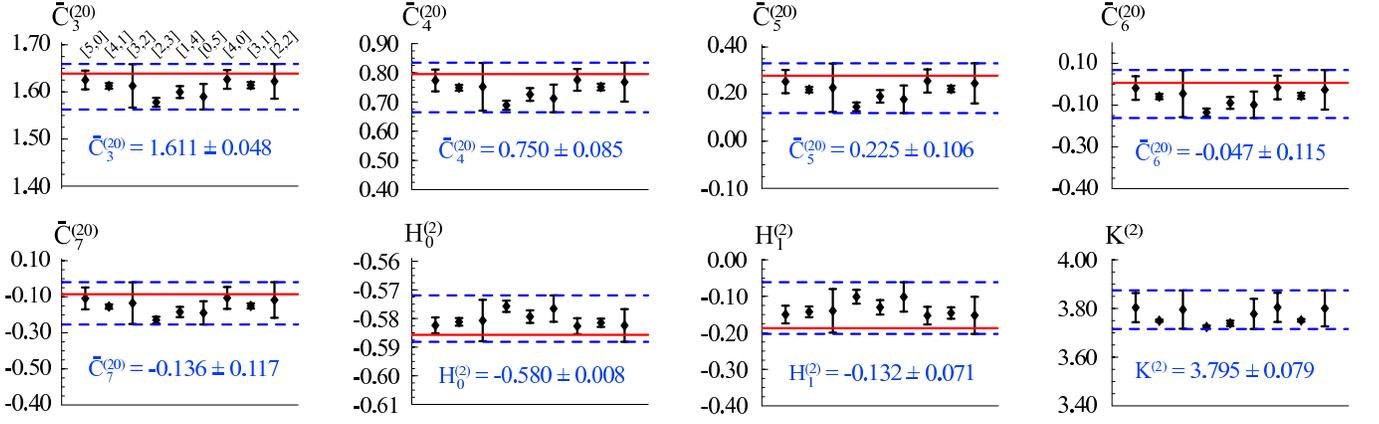}
\caption{Results from the reconstructed $\Pi^{(2)}$ function in
  approximation~C for the
  coefficients $C^{(20)}_{3,4,5,6,7}$ that arise in the expansion around
  $z=0$, for the coefficients $H^{(2)}_{0,1}$ that occur in the
  non-logarithmic terms in the high-energy expansion $|z|\to\infty$ and
  for the constant $K^{(2)}$ that appears in the expansion at threshold around
  $z=1$. The red solid lines represent the respective exact results known from 
  computation of Feynman diagrams. The dashed blue lines represent the
  envelope of all results obtained from the reconstructed $\Pi^{(2)}$
  functions. The individual error bars represent the range of vales obtained
  from the reconstructed $\Pi^{(2)}$ functions using one particular Pad\'e
  approximant $P_{m,n}$. The various types of Pad\'e approximants that have
  been used are indicated in the upper left panel; the same order is used for
  all other panels.    
  All results are for $n_f=n_\ell+1=4$ running flavors relevant for
  charm production.   
\label{fig:2momOas2}
}
\end{center}
\end{figure}

Given the analytic form for the reconstructed $\Pi^{(2)}$
functions we can expand them in the threshold region, the high-energy limit
and around $q^2=0$. 
In Fig.~\ref{fig:2momOas2} the results for the coefficients $\bar C^{(20)}_k$
for $k=3,4,5,6,7$, the high-energy constants $H^{(2)}_{0,1}$ and the
threshold constant $K^{(2)}$ are displayed exemplarily for
$n_f=n_\ell+1=4$ relevant for the production of charm quarks. The labels
$[m,n]$ which have been   
added to the upper left panel for $\bar C^{(20)}_3$ refer to the
Pad\'e approximant used for the respective $\Pi^{(2)}_{\rm reg}$ functions, and
their order is representative for all diagrams. 
The error bars represent the range of values covered by the variations of the
modification factors as described in Sec.~\ref{sectionPilog}. The blue dashed lines
indicate the range covered by all individual results and the red solid lines
show the exact result obtained from Feynman diagrams. We see that for all
cases the exact values are well within the range covered by the reconstructed
$\Pi^{(2)}$ functions. Particularly precise determination are obtained for
$\bar C^{(20)}_3$ and the leading high-energy coefficient $H^{(2)}_0$. We
also obtain a very precise determination of the the threshold constant
$K^{(2)}$.

\begin{table}[ht]
\begin{center}
\begin{tabular}{|r|r|r|r|r|r|r|}
\hline
&  approx. A\mbox{\quad} & approx. B\mbox{\quad} & 
approx. C\mbox{\quad} & approx. D\mbox{\quad}  & 
approx. E\mbox{\quad}  & exact\mbox{\,\,} 
\\ \hline
$\bar C^{(20)}_1$ &&&&&& $2.49671$ 
\\ \hline
$\bar C^{(20)}_2$ &&&&&& $2.77702$
\\ \hline
$\bar C^{(20)}_3$ 
& $ 1.365\pm 0.425 $ & $ 1.609\pm 0.266 $ & $ 1.611\pm 0.048 $ & & & $ 1.63882 $
\\ \hline
$\bar C^{(20)}_4$ 
& $ 0.283\pm 0.799 $ & $ 0.770\pm 0.441 $ & $ 0.750\pm 0.085 $ & & & $ 0.79555 $
\\ \hline
$\bar C^{(20)}_5$
& $ - 0.389\pm 1.057 $ & $ 0.271\pm 0.521 $ & $ 0.225\pm 0.106 $ & $ 0.278\pm 
0.001 $ & & $ 0.27814 $
\\ \hline
$\bar C^{(20)}_6$
& $ - 0.744\pm 1.213 $ & $ 0.021\pm 0.541 $ & $ - 
            0.047\pm 0.115 $ & $ 0.007\pm 0.002 $ & & $ 0.0070080 $
\\ \hline
$\bar C^{(20)}_7$
& $ - 0.871\pm 1.296 $ & $ - 0.054\pm 0.528 $ & $ - 0.136\pm 0.117 $ & $ - 
              0.086\pm 0.003 $ & $ - 0.08594\pm 0.00003 $ & $ - 0.085963 $
\\ \hline
$H^{(2)}_0$
& $ 0.159\pm 0.770 $ & $ - 0.561\pm 0.063 $ & $ - 0.580\pm 0.008 $ & $ - 
              0.5854\pm 0.0004 $ & $ - 0.5857\pm 0.0001 $ & $ - 0.58570 $
\\ \hline
$H^{(2)}_1$
& $ 0.007\pm 0.574 $ & $ 0.338\pm 0.871 $ & $ - 0.132\pm 0.071 $ & $ - 
              0.180\pm 0.008 $ & $ - 0.185\pm 0.004 $ & $ - 0.18723 $
\\ \hline
$K^{(2)}$
& $ - 6.64\pm 10.12 $ & $ 3.933\pm 0.303 $ & $ 3.795\pm 0.079 $ & $ \
3.809\pm 0.032 $ & $ 3.805\pm 0.020 $ &
\\ \hline
\end{tabular}
\caption{
Results for the coefficients $\bar C^{(20)}_{3,4,5,6,7}$, 
$H^{(2)}_{0,1}$ and $K^{(2)}$ from the reconstructed $\Pi^{(2)}$ functions
using various different types of approximations for $\Pi^{(2)}$.    
Empty entries for coefficients $\bar C^{(20)}_k$ indicate that they are exact
in that particular approximation. The exact analytic form for $K^{(2)}$ is
unknown. 
All results are for $n_f=n_\ell+1=4$ running flavors relevant for
charm production.
\label{tab:CHKOas2}
}
\end{center}
\end{table}

To demonstrate that our approach is systematic we need to show that the
results become more accurate once more information is included for the
reconstruction of $\Pi^{(2)}$. In Tab.~\ref{tab:CHKOas2} the results of
Fig.~\ref{fig:2momOas2} for $\bar C^{(20)}_k$ with $k=3,4,5,6,7$,
$H^{(2)}_{0,1}$ and $K^{(2)}$ are displayed in the line labeled as
``approximation~C''. In comparison we also show the results when all the
information $\sim 1/z^2$ in the high-energy expansion is neglected for the
reconstruction of $\Pi^{(2)}$ (``approximation~B'') and when in addition to
that also all NNLO threshold information is neglected (``approximation~A''). 
The results show that the properties of the vacuum polarization can be
determined more accurately once more information is used for its
reconstruction with our approach. Moreover, we find that the variations of the
reconstructed vacuum polarization function due to the different choices for the
modification factors and the Pad\'e approximants represent a reliable tool to
estimate the uncertainties.

\begin{figure}[ht]
\begin{center}
\epsfxsize= 9cm
\epsffile{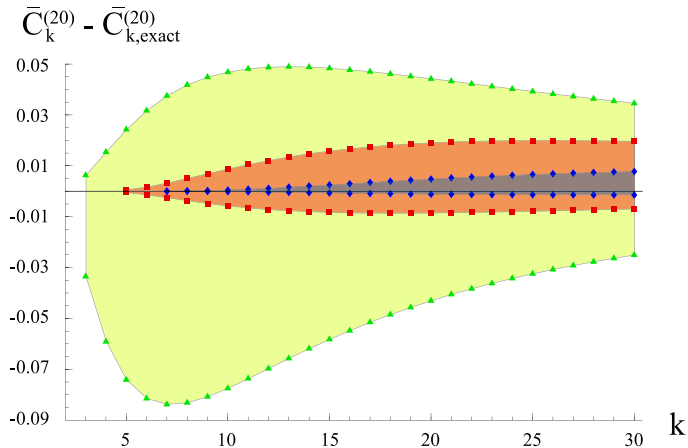}
\caption{ 
Difference of the values for $\bar C^{(20)}_k$, determined from the
reconstructed $\Pi^{(2)}$ functions based on Taylor-like Pad\'e approximants, 
and the known exact values $\bar C^{(20)}_{k,\rm exact}$. The results are
shown for approximation~C (green triangles and light greed
shaded area), D~(red squared symbols and medium red shaded area) and 
E~(blue diamonds and dark blue shaded area).
The shaded areas represent the variation of the results due to the changes in
the modification factors. 
\label{fig:Cnhigh}
}
\end{center}
\end{figure}

It is an astounding and amusing fact our approach allows for a determination of the
coefficients $\bar C^{(20)}_k$ for large values of $k$ with practically
negligible uncertainties. This is demonstrated in Fig.~\ref{fig:Cnhigh} where
the difference of the results we obtain for the coefficients $\bar
C^{(20)}_k$ and the exact values from Ref.~\cite{Maier:2007yn,Boughezal:2006uu}, 
$\bar C^{(20)}_k-\bar C^{(20)}_{k,{\rm exact}}$
are shown up to $k=30$. The range of values between the green triangular-shaped
symbols (green light shaded region) is obtained from approximation~C using
solutions with Taylor-like Pad\'e approximants. For $k>(10,15,20)$ the maximal
relative discrepancy to the exact values is below $(14,3,1)$\%. The
discrepancies become even smaller when more of the coefficients in the
expansion around $z=0$ of Eq.~(\ref{Pismallz}) are accounted for in the
reconstruction of $\Pi^{(2)}$. Including the coefficients up to order $z^4$
(approximation~D) we obtain the range of values between the
red squared symbols (red shaded region). In this
case we obtain for $k>(10,15,20)$ a maximal relative discrepancy to the exact
values of below $(3,0.8,0.5)$\%. Including the coefficients up to order $z^6$
(approximation~E) we obtain
the range of values between the blue diamond-shaped symbols (blue dark shaded
region). Here, we obtain for $k>10$ a maximal relative discrepancy
to the exact values of below $0.2$\%. 
The results for the high-energy coefficients $H^{(2)}_{0,1}$ for
approximations~D and E are also shown in Tab.~\ref{tab:CHKOas2}. Again we find
agreement with the exact results with decreasing uncertainties once more
information is included for the reconstruction of $\Pi^{(2)}$.
Given the excellent quality of the results we consider our approach a reliable method
to determine the previously unknown threshold constant $K^{(2)}$. As our final result for
$K^{(2)}$ we adopt
\begin{align}
K^{(2)} \, = \, 3.81 \, \pm \, 0.02
\,.
\end{align}

\begin{figure}[t]
\begin{center}
\epsfxsize=\textwidth
\epsffile{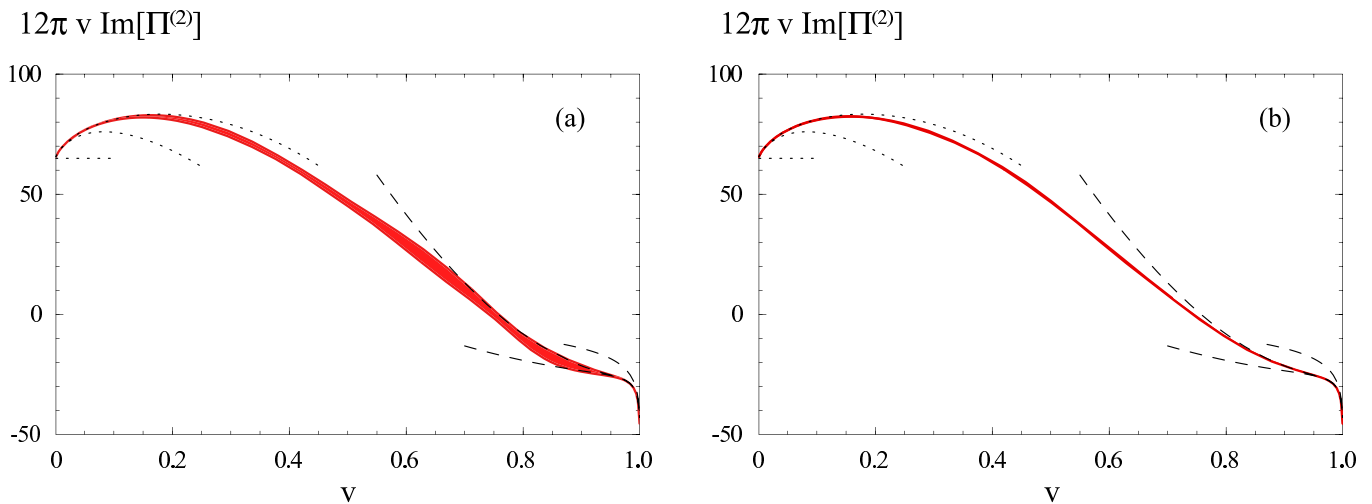}
\caption{ 
Results for  $12 \pi v \mbox{Im}[\Pi^{(2)}(q^2+i 0)]$ as a function of $v$ for
$n_f=4$. The red bands represent the uncertainties. In the left panel the
results are based on the reconstructed 
$\Pi^{(2)}$ function incorporating the coefficients in the small-$z$ expansion up to
order $z^2$ (approximation~C) and in the right panel the coefficients up to
order $z^6$ are accounted for. The dotted and dashed black lines show the 
expansions in the threshold and the high-energy region up to NNLO.
See the text for more  details. 
\label{fig:ROas2}
}
\end{center}
\end{figure}

To conclude this section let us analyze the ${\cal O}(\alpha_s^2)$ corrections
to the $e^+e^-$ cross section obtained from the reconstructed $\Pi^{(2)}$. 
In Fig.~\ref{fig:ROas2} we have
plotted $12 \pi v \mbox{Im}[\Pi^{(2)}(q^2+i 0)]$ for $n_f=4$ relevant for charm
quark production in the pole mass scheme as a function of
the quark velocity $v=\sqrt{1-1/z}$. We have included the factor of $v$ to
suppress the Coulomb $1/v$-singularity that arises in the cross
section for small values of $v$ and to have a finite value in the limit $v\to
0$. In the left panel the result for 
approximation~C is shown. The red band is the area
covered by all solutions for $\Pi^{(2)}$ that pass the criteria
discussed in Sec.~\ref{sectionunphysical} and represents the uncertainty.  
The size of the uncertainty corresponds to the envelope of the individual
error bars shown in Fig.~\ref{fig:2momOas2} where
approximation~C has been used as well. For comparison we have also displayed
the expansions in the threshold region for $v\to 0$ (dotted lines) and in the
high-energy limit for $v\to 1$ (dashed lines), where the short lines refer to
leading order, the medium-length lines to next-to-leading order and the
longest lines to next-to-next-to-leading order.  
The uncertainties are reduced substantially when additional coefficients for
the expansion around $z=0$ are 
included for the reconstruction of $\Pi^{(2)}$. This is demonstrated in the
right panel, where the coefficients up to order $z^6$ are included for
the reconstruction 
of $\Pi^{(2)}$. Again the red band is the area covered by all solutions for
$\Pi^{(2)}$ that pass the criteria discussed in
Sec.~\ref{sectionunphysical}. For method~(i) to account the Coulomb
singularity we found solutions based on the Pad\'e approximants [9,0], [8,1],
[7,2], [6,3], [5,4], [3,6], [1,8], and for method~(ii) we found solutions
based on the Pad\'e approximants  [8,0], [7,1], [6,2], [5,3], [4,4], [3,5],
[1,7]. The width of the band is already smaller than the width of the solid lines
used to draw the boundaries of the band. For $v=(0.2,0.4,0.6,0.8)$ the relative
uncertainty 
is $\pm(0.09,0.4,2.0,2.5)\%$ and thus negligible for all conceivable
practical applications. The approximation formulae for the 
${\cal O}(\alpha_s^2 C_F^2)$ and  ${\cal O}(\alpha_s^2 C_A C_F)$ contributions 
given in  Ref.~\cite{Chetyrkin:1996cf,Chetyrkin:1997mb} (using Eqs.~(65) and
(66) of Ref.~\cite{Chetyrkin:1996cf}) together with the analytically known
fermionic corrections agree within 1-2\% with our result. We thus 
confirm the results for the cross section given in
Refs.~\cite{Chetyrkin:1996cf,Chetyrkin:1997mb}.

\section{Analysis for the Vacuum Polarization at ${\cal O}(\alpha_s^3)$}
\label{sectionOas3}

For the reconstruction of $\Pi^{(3)}$ we use all available information from
the expansions in the threshold region, Eqs.~(\ref{Pithresh}), the high-energy
region, Eqs.~(\ref{Pihigh}) and around $z=0$ in Eqs.~(\ref{Pismallz}). For the
construction of $\Pi^{(3)}_{\rm reg}$ we account for the first two
coefficients in the expansion around $z=0$, the non-logarithmic term
$\propto\sqrt{1-z}$ in the threshold limit and the two constraints from the
absence of terms $\sim 1/z^{3/2}$ and $\sim 1/z^{5/2}$ for $|z|\to\infty$.
This amounts to~6 constraints on the Pad\'e approximants for method (i), where
the Coulomb singularity $\propto 1/(1-z)$ is accounted for in $\Pi^{(3)}_{\rm
reg}$, and 5 constraints on the Pad\'e approximants for method (ii), where
this Coulomb singularity is accounted for in $\Pi^{(3)}_{\rm log}$. Thus we
have $n+m=5$ for the Pad\'e approximants $P_{m,n}$ for method (i) and $n+m=4$
for the Pad\'e approximants $P_{m,n}$ for method (ii).

\begin{figure}[ht]
\begin{center}
\epsfxsize=\textwidth
\epsffile{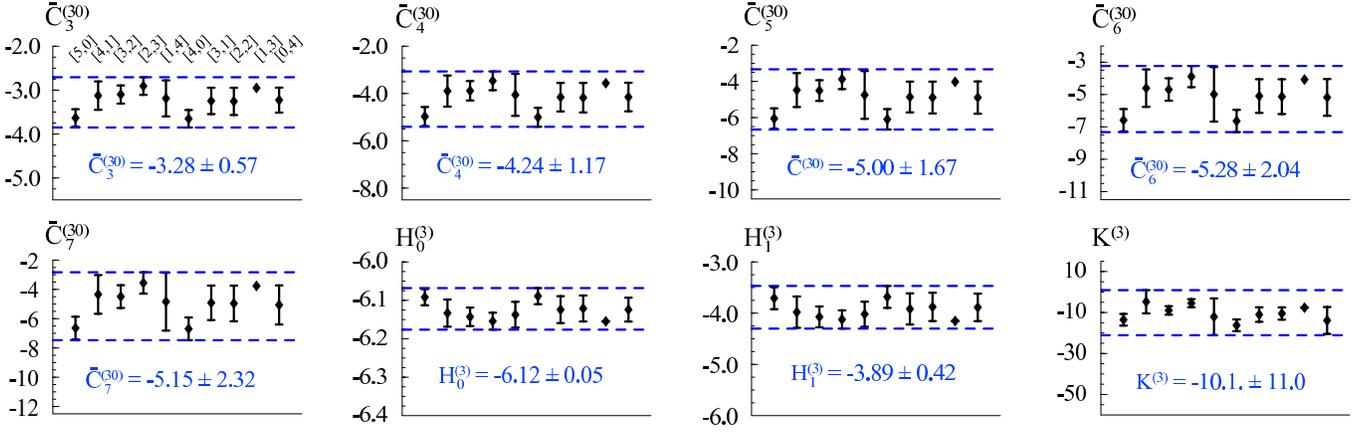}
\caption{ Results from the reconstructed $\Pi^{(3)}$ function for the
  coefficients $C^{(30)}_{3,4,5,6,7}$ that arise in the expansion around
  $z=0$, for the coefficients $H^{(3)}_{0,1}$ that occur in the
  non-logarithmic terms in the high-energy expansion $|z|\to\infty$ and
  for the constant $K^{(3)}$ that appears in the expansion at threshold around
  $z=1$. The dashed blue lines represent the
  envelope of all results obtained from the reconstructed $\Pi^{(3)}$
  functions. The individual error bars represent the range of vales obtained
  from the reconstructed $\Pi^{(3)}$ functions using one particular Pad\'e
  approximant $P_{m,n}$. The various types of Pad\'e approximants that have
  been used are indicated in the upper left panel; the same order is used for
  all other panels.    
  All results are for $n_f=n_\ell+1=4$ running flavors relevant for
  charm production.   
\label{fig:2momOas3}
}
\end{center}
\end{figure}

In Fig.~\ref{fig:2momOas3} the results for the coefficients $\bar C^{(30)}_k$
for $k=3,4,5,6,7$, the high-energy constants $H^{(3)}_{0,1}$ and the
threshold constant $K^{(3)}$ are displayed for $n_f=n_\ell+1=4$ relevant for
charm quark production. The different labels $[m,n]$ which 
have been  added to the upper left panel for $\bar C^{(30)}_3$ refer to the
Pad\'e approximant used for the respective $\Pi^{(3)}_{\rm reg}$ function and
their order is representative for all panels. The error bars represent the
range of values covered by the variations of the 
modification factors as described in Sec.~\ref{sectionPilog}, and the blue
dashed lines indicate the range covered by all individual results. We adopt
this range as the uncertainty in our determination of these coefficients, and
the results are summarized together with the corresponding results for
$n_f=n_\ell+1=5$ in Tab.~\ref{tab:CHKOas3}. For the determination of the 
high-energy coefficients 
$H^{(3)}_0$ and $H^{(3)}_1$ we find uncertainties of about 1\% and 10\%,
respectively. This compares well with the corresponding results for
$H^{(2)}_0$ and $H^{(2)}_1$ we have obtained at ${\cal O}(\alpha_s^2)$ for
approximation~C, see Fig.~\ref{fig:2momOas2} and Tab.~\ref{tab:CHKOas2}. 
For the coefficients $\bar C^{(30)}_k$ with $k\ge 3$ we find 
somewhat larger relative uncertainties than in for the $\bar C^{(20)}_k$ in
approximation~C. This is, however, not  
unexpected since the cancellations that arise when the pole mass results for
these coefficients are
transferred to the $\overline{\mbox{MS}}$ mass scheme are substantially larger
at ${\cal O}(\alpha_s^3)$. 
The result for $K^{(3)}$ has a particularly large error
and can merely serve as a rough constraint on its true values. Concerning
the precision in the determinations of $K^{(3)}$, we believe that a substantial
improvement can be achieved once the full set of NNNLO 
terms $\propto \sqrt{1-z}$ in the expansion for $R$ at the threshold  
and the exact values for $\bar C^{(30)}_k$ with $k\ge 3$ become available.

\begin{table}[ht]
\begin{center}
\begin{tabular}{|c|c|c|}
\hline
& $n_f=4$ & $n_f=5$ 
\\ \hline
$\bar C^{(30)}_1$ &
$-5.6404$ & $-7.7624$
\\ \hline
$\bar C^{(30)}_2$ & 
$-3.4937$ & $-2.6438$
\\ \hline
$\bar C^{(30)}_3$ &
$-3.279 \pm 0.573$ & $-1.457 \pm 0.579$
\\ \hline
$\bar C^{(30)}_4$  &
$-4.238 \pm 1.171$ & $-1.935 \pm 1.201$
\\ \hline
$\bar C^{(30)}_5$ &
$-4.996 \pm 1.666$ & $-2.507 \pm 1.732$
\\ \hline
$\bar C^{(30)}_6$ &
$-5.280 \pm 2.045$ & $-2.809 \pm 2.150$
\\ \hline
$\bar C^{(30)}_7$ &
$-5.151 \pm 2.321$ & $-2.847 \pm 2.467$
\\ \hline
$H^{(3)}_0$ &
$-6.122 \pm 0.054$ & $-4.989 \pm 0.053$
\\ \hline
$H^{(3)}_1$ &
$-3.885 \pm 0.417$ & $-3.180 \pm 0.405$
\\ \hline
$K^{(3)}$ &
$-10.09 \pm 11.00$ & $-5.97 \pm 10.09$
\\ \hline
\end{tabular}
\caption{Summary of the results for the
  coefficients $C^{(30)}_{3,4,5,6,7}$, $H^{(3)}_{0,1}$ and $K^{(3)}$ obtained
  from the reconstructed $\Pi^{(3)}$ function for $n_f=n_\ell+1=4$ and
  $n_f=n_\ell+1=5$. The coefficients $C^{(30)}_{1,2}$ are known exactly and
  shown for completeness.
\label{tab:CHKOas3}
}
\end{center}
\end{table}

One of the most important applications of the coefficients $\bar C^{(30)}_n$
is the determination of the $\overline{\mbox{MS}}$ charm and
bottom quark masses from moments $M_n$ of the charm and bottom quark $e^+e^-$
cross section. For small values of $n$ one way to compute the moments is using
fixed-order perturbation theory as shown in Eq.~(\ref{Mnexpanded}). Using the
results from Tab.~\ref{tab:CHKOas3} we find for the fixed-order moments at
${\cal O}(\alpha_s^3)$ for charm quarks ($n_f=4$) 
\begin{align}
M_3\, =\,&(0.1348\pm 0.0044\pm 0.0005)\times 10^{-2}\,, 
\nn\\[1mm] 
M_4\, = \,& (0.153\pm 0.032\pm 0.002)\times 10^{-3}\,,
\nn\\[1mm] 
M_5\, = \,& (0.199\pm 0.084\pm 0.008)\times 10^{-4}\,
\nn\\[1mm] 
M_6\, = \,& (0.084\pm 0.144\pm 0.036)\times 10^{-5}\,.
\end{align}
Here we used $\overline m_c(\overline m_c)=1.27$~GeV for
the $\overline{\mbox{MS}}$ charm mass and
$\alpha_s^{(n_f=4)}(1.27~\mbox{GeV})=0.387637$ for the strong coupling as the
input and four-loop renormalization group evolution. 
The first error arises from the variation of the renormalization scale
between $1.27$ and $3.81$~GeV and the second error is due to the uncertainties
in the ${\cal O}(\alpha_s^3)$ coefficients $C^{(30)}_k$ shown in
Tab.~\ref{tab:CHKOas3}.  
For bottom quarks ($n_f=5$) with 
$\overline m_b(\overline m_b)=4.17$~GeV and
$\alpha_s^{(n_f=5)}(4.17~\mbox{GeV})=0.224778$ as the input we find
\begin{align}
M_3\, = \, & (2.350\pm 0.017\pm 0.002)\times 10^{-7}\,, 
\nn\\[1mm] 
M_4\, = \, &(2.167\pm 0.045\pm 0.005)\times 10^{-9} \,,  
\nn\\[1mm] 
M_5\, = \, &(2.126\pm 0.091\pm 0.011)\times 10^{-11}\,, 
\nn\\[1mm] 
M_6\, = \, &(2.160\pm 0.148\pm 0.022)\times 10^{-13}\,.
\end{align}
The first error arises from the variation of the renormalization scale
between $2.085$ and $8.34$~GeV and the second error is due to uncertainties in the
${\cal O}(\alpha_s^3)$ coefficients $C^{(30)}_k$ shown in
Tab.~\ref{tab:CHKOas3}.  
The results show that that the uncertainties in $M_{3,4,5}$ caused by the 
errors in the coefficients $C^{(30)}_{3,4,5}$ we have obtained in this work 
are an order of magnitude smaller than the overall uncertainties of the
moments at ${\cal O}(\alpha_s^3)$ due to variations of the renormalization
scale. For physically relevant values of $n$ they can be safely neglected.

\begin{figure}[t]
\begin{center}
\epsfxsize= 9cm
\epsffile{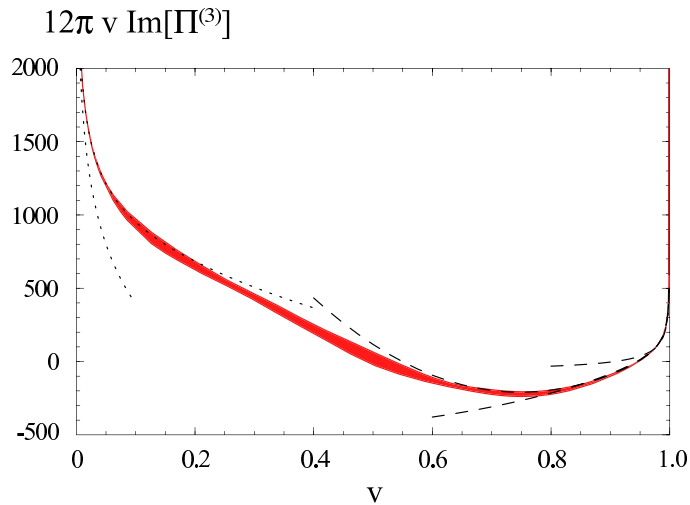}
\caption{ 
Result for  $12 \pi v \mbox{Im}[\Pi^{(3)}(q^2+i 0)]$ as a function of $v$ for
$n_f=4$ using the currently available information for the reconstruction of
$\Pi^{(3)}$.The red band represent the uncertainty. The dotted and dashed
black lines show the expansions in the threshold and the high-energy region up
to next-to-next-to-leading order. See the text for details. 
\label{fig:ROas3}
}
\end{center}
\end{figure}

Finally, let us analyze the ${\cal O}(\alpha_s^3)$ corrections to the $e^+e^-$
cross section obtained from the $\Pi^{(3)}$. In Fig.~\ref{fig:ROas3} we have
plotted the function $12 \pi v \mbox{Im}[\Pi^{(3)}(q^2+i 0)]$ for $n_f=4$
relevant for charm quark production in the pole mass scheme as a function of 
the quark velocity $v=\sqrt{1-1/z}$. As for the analysis in
Fig.~\ref{fig:ROas2} we have included the factor $v$ to suppress the Coulomb
singularity. The function still diverges 
logarithmically for $v\to 0$ because the ${\cal O}(\alpha_s^3)$ cross section  
has a singularity $\sim \ln(v)/v$ in the nonrelativistic limit.
The red shaded band is the area covered by all solutions for $\Pi^{(3)}$ 
that pass the criteria discussed in
Sec.~\ref{sectionunphysical} and represents the uncertainty. 
The relative uncertainty is about 10\% at $v=0.2$ and $0.8$ and should be
acceptable for most applications where ${\cal O}(\alpha_s^3)$ accuracy is
required. For comparison we have also displayed
the expansions in the threshold region for $v\to 0$ (dotted lines) at NLO
(short line) and at NNLO (long line). Likewise the expansions in the
high-energy limit for $v\to 1$ (dashed lines) are shown, where the short line
refers to order $1/z^0$, the medium-length line to order $1/z$ and the longest
lines to order $1/z^2$.  
We strongly emphasize the importance of incorporating the
NNLO contributions in the expansion close to the threshold and the $1/z^2$
terms at high energies for achieving our result. Once more information from
the different kinematic regions becomes available, the uncertainties can be
further reduced substantially.

\section{Conclusions}
\label{sectionconclusions}

In this work we have determined the full mass and $q^2$ dependence of the
${\cal O}(\alpha_s^2)$ and ${\cal O}(\alpha_s^3)$ corrections to the heavy
quark vacuum polarization function $\Pi(q^2,m^2)$ and its contribution to the
$e^+e^-$ total cross section. 
Our approach uses known results for the expansions of $\Pi(q^2,m^2)$ at high
energies, in the threshold region and around $q^2=0$, conformal mapping and
the Pad\'e approximation method. We have demonstrated for the vacuum
polarization function at ${\cal O}(\alpha_s^2)$ that the approach allows for
reliable determinations of other properties of $\Pi$ with small uncertainties, and
that the uncertainties of the results can be systematically reduced if more
information from the three different kinematic regions is accounted for. Our
results for the cross section at ${\cal O}(\alpha_s^2)$ also confirm previous
results by Chetyrkin, K\"uhn and Steinhauser from
Refs.~\cite{Chetyrkin:1996cf,Chetyrkin:1997mb}. For 
the vacuum polarization function at ${\cal O}(\alpha_s^2)$ we have determined
the previously unknown non-logarithmic constant term that arises at NLO in the 
expansion close to the threshold. For the ${\cal O}(\alpha_s^3)$ corrections
to the vacuum polarization function we determined the previously unknown
coefficients in the expansion around $q^2=0$ beyond order $q^4$ and the
first two non-logarithmic coefficients in the high-energy expansion. The
results for the coefficients in the expansion around $q^2=0$ allow for the
determination of the moments $M_n$ of the $e^+e^-$ cross section for $n\ge 3$
at ${\cal O}(\alpha_s^3)$. 

\vskip 5mm
\section{Acknowledgements}
\label{sectionacknowledgements}

We thank K.~Chetyrkin, H.~K\"uhn and M.~Steinhauser for useful converstion and
comments to the manuscript.
M.~Zebarjad thanks the MPI for hospitality while this work was 
accomplished and the MPI guest program for partial support.  
This work was supported in part by
the EU network contract MRTN-CT-2006-035482 (FLAVIAnet).

\vskip 7mm
\noindent
{\bf Note added:} After completion of this work K.~Chetyrkin pointed 
out to us that analytic expressions for the constants $H^{(3)}_1$ can be
derived from results given in Ref.~\cite{Baikov:2004ku}. 
Evaluated numerically they give $H_1^{(3)}=-4.33306$ for $n_f=4$ and  
$H_1^{(3)}=-3.53165$ for $n_f=5$ which is in agreement with the results we
have presented in Tab.~\ref{tab:CHKOas3}. 


\bibliography{vacpolpaper}

\begin{thebibliography}{32}
\expandafter\ifx\csname natexlab\endcsname\relax\def\natexlab#1{#1}\fi
\expandafter\ifx\csname bibnamefont\endcsname\relax
  \def\bibnamefont#1{#1}\fi
\expandafter\ifx\csname bibfnamefont\endcsname\relax
  \def\bibfnamefont#1{#1}\fi
\expandafter\ifx\csname citenamefont\endcsname\relax
  \def\citenamefont#1{#1}\fi
\expandafter\ifx\csname url\endcsname\relax
  \def\url#1{\texttt{#1}}\fi
\expandafter\ifx\csname urlprefix\endcsname\relax\def\urlprefix{URL }\fi
\providecommand{\bibinfo}[2]{#2}
\providecommand{\eprint}[2][]{\url{#2}}

\bibitem[{\citenamefont{Novikov et~al.}(1978)}]{Novikov:1977dq}
\bibinfo{author}{\bibfnamefont{V.~A.} \bibnamefont{Novikov}}
  \bibnamefont{et~al.}, \bibinfo{journal}{Phys. Rept.}
  \textbf{\bibinfo{volume}{41}}, \bibinfo{pages}{1} (\bibinfo{year}{1978}).

\bibitem[{\citenamefont{Reinders et~al.}(1985)\citenamefont{Reinders,
  Rubinstein, and Yazaki}}]{Reinders:1984sr}
\bibinfo{author}{\bibfnamefont{L.~J.} \bibnamefont{Reinders}},
  \bibinfo{author}{\bibfnamefont{H.}~\bibnamefont{Rubinstein}},
  \bibnamefont{and} \bibinfo{author}{\bibfnamefont{S.}~\bibnamefont{Yazaki}},
  \bibinfo{journal}{Phys. Rept.} \textbf{\bibinfo{volume}{127}},
  \bibinfo{pages}{1} (\bibinfo{year}{1985}).

\bibitem[{\citenamefont{Kallen and Sabry}(1955)}]{Kallen:1955fb}
\bibinfo{author}{\bibfnamefont{A.~O.~G.} \bibnamefont{Kallen}}
  \bibnamefont{and} \bibinfo{author}{\bibfnamefont{A.}~\bibnamefont{Sabry}},
  \bibinfo{journal}{Kong. Dan. Vid. Sel. Mat. Fys. Med.}
  \textbf{\bibinfo{volume}{29N17}}, \bibinfo{pages}{1} (\bibinfo{year}{1955}).

\bibitem[{\citenamefont{Kniehl}(1990)}]{Kniehl:1989kz}
\bibinfo{author}{\bibfnamefont{B.~A.} \bibnamefont{Kniehl}},
  \bibinfo{journal}{Phys. Lett.} \textbf{\bibinfo{volume}{B237}},
  \bibinfo{pages}{127} (\bibinfo{year}{1990}).

\bibitem[{\citenamefont{Hoang et~al.}(1994)\citenamefont{Hoang, Jezabek, Kuhn,
  and Teubner}}]{Hoang:1994it}
\bibinfo{author}{\bibfnamefont{A.~H.} \bibnamefont{Hoang}},
  \bibinfo{author}{\bibfnamefont{M.}~\bibnamefont{Jezabek}},
  \bibinfo{author}{\bibfnamefont{J.~H.} \bibnamefont{Kuhn}}, \bibnamefont{and}
  \bibinfo{author}{\bibfnamefont{T.}~\bibnamefont{Teubner}},
  \bibinfo{journal}{Phys. Lett.} \textbf{\bibinfo{volume}{B338}},
  \bibinfo{pages}{330} (\bibinfo{year}{1994}), \eprint{hep-ph/9407338}.

\bibitem[{\citenamefont{Hoang et~al.}(1995)\citenamefont{Hoang, Kuhn, and
  Teubner}}]{Hoang:1995ex}
\bibinfo{author}{\bibfnamefont{A.~H.} \bibnamefont{Hoang}},
  \bibinfo{author}{\bibfnamefont{J.~H.} \bibnamefont{Kuhn}}, \bibnamefont{and}
  \bibinfo{author}{\bibfnamefont{T.}~\bibnamefont{Teubner}},
  \bibinfo{journal}{Nucl. Phys.} \textbf{\bibinfo{volume}{B452}},
  \bibinfo{pages}{173} (\bibinfo{year}{1995}), \eprint{hep-ph/9505262}.

\bibitem[{\citenamefont{Chetyrkin
  et~al.}(1996{\natexlab{a}})\citenamefont{Chetyrkin, Kuhn, and
  Steinhauser}}]{Chetyrkin:1995ii}
\bibinfo{author}{\bibfnamefont{K.~G.} \bibnamefont{Chetyrkin}},
  \bibinfo{author}{\bibfnamefont{J.~H.} \bibnamefont{Kuhn}}, \bibnamefont{and}
  \bibinfo{author}{\bibfnamefont{M.}~\bibnamefont{Steinhauser}},
  \bibinfo{journal}{Phys. Lett.} \textbf{\bibinfo{volume}{B371}},
  \bibinfo{pages}{93} (\bibinfo{year}{1996}{\natexlab{a}}),
  \eprint{hep-ph/9511430}.

\bibitem[{\citenamefont{Chetyrkin
  et~al.}(1996{\natexlab{b}})\citenamefont{Chetyrkin, Kuhn, and
  Steinhauser}}]{Chetyrkin:1996cf}
\bibinfo{author}{\bibfnamefont{K.~G.} \bibnamefont{Chetyrkin}},
  \bibinfo{author}{\bibfnamefont{J.~H.} \bibnamefont{Kuhn}}, \bibnamefont{and}
  \bibinfo{author}{\bibfnamefont{M.}~\bibnamefont{Steinhauser}},
  \bibinfo{journal}{Nucl. Phys.} \textbf{\bibinfo{volume}{B482}},
  \bibinfo{pages}{213} (\bibinfo{year}{1996}{\natexlab{b}}),
  \eprint{hep-ph/9606230}.

\bibitem[{\citenamefont{Czakon and Schutzmeier}(2008)}]{Czakon:2007qi}
\bibinfo{author}{\bibfnamefont{M.}~\bibnamefont{Czakon}} \bibnamefont{and}
  \bibinfo{author}{\bibfnamefont{T.}~\bibnamefont{Schutzmeier}},
  \bibinfo{journal}{JHEP} \textbf{\bibinfo{volume}{07}}, \bibinfo{pages}{001}
  (\bibinfo{year}{2008}), \eprint{0712.2762}.

\bibitem[{\citenamefont{Chetyrkin et~al.}(2000)\citenamefont{Chetyrkin,
  Harlander, and Kuhn}}]{Chetyrkin:2000zk}
\bibinfo{author}{\bibfnamefont{K.~G.} \bibnamefont{Chetyrkin}},
  \bibinfo{author}{\bibfnamefont{R.~V.} \bibnamefont{Harlander}},
  \bibnamefont{and} \bibinfo{author}{\bibfnamefont{J.~H.} \bibnamefont{Kuhn}},
  \bibinfo{journal}{Nucl. Phys.} \textbf{\bibinfo{volume}{B586}},
  \bibinfo{pages}{56} (\bibinfo{year}{2000}), \eprint{hep-ph/0005139}.

\bibitem[{\citenamefont{Harlander and Steinhauser}(2003)}]{Harlander:2002ur}
\bibinfo{author}{\bibfnamefont{R.~V.} \bibnamefont{Harlander}}
  \bibnamefont{and}
  \bibinfo{author}{\bibfnamefont{M.}~\bibnamefont{Steinhauser}},
  \bibinfo{journal}{Comput. Phys. Commun.} \textbf{\bibinfo{volume}{153}},
  \bibinfo{pages}{244} (\bibinfo{year}{2003}), \eprint{hep-ph/0212294}.

\bibitem[{\citenamefont{Hoang and Teubner}(1998)}]{Hoang:1998xf}
\bibinfo{author}{\bibfnamefont{A.~H.} \bibnamefont{Hoang}} \bibnamefont{and}
  \bibinfo{author}{\bibfnamefont{T.}~\bibnamefont{Teubner}},
  \bibinfo{journal}{Phys. Rev.} \textbf{\bibinfo{volume}{D58}},
  \bibinfo{pages}{114023} (\bibinfo{year}{1998}), \eprint{hep-ph/9801397}.

\bibitem[{\citenamefont{Hoang}(1997)}]{Hoang:1997sj}
\bibinfo{author}{\bibfnamefont{A.~H.} \bibnamefont{Hoang}},
  \bibinfo{journal}{Phys. Rev.} \textbf{\bibinfo{volume}{D56}},
  \bibinfo{pages}{7276} (\bibinfo{year}{1997}), \eprint{hep-ph/9703404}.

\bibitem[{\citenamefont{Chetyrkin et~al.}(2006)\citenamefont{Chetyrkin, Kuhn,
  and Sturm}}]{Chetyrkin:2006xg}
\bibinfo{author}{\bibfnamefont{K.~G.} \bibnamefont{Chetyrkin}},
  \bibinfo{author}{\bibfnamefont{J.~H.} \bibnamefont{Kuhn}}, \bibnamefont{and}
  \bibinfo{author}{\bibfnamefont{C.}~\bibnamefont{Sturm}},
  \bibinfo{journal}{Eur. Phys. J.} \textbf{\bibinfo{volume}{C48}},
  \bibinfo{pages}{107} (\bibinfo{year}{2006}), \eprint{hep-ph/0604234}.

\bibitem[{\citenamefont{Boughezal
  et~al.}(2006{\natexlab{a}})\citenamefont{Boughezal, Czakon, and
  Schutzmeier}}]{Boughezal:2006px}
\bibinfo{author}{\bibfnamefont{R.}~\bibnamefont{Boughezal}},
  \bibinfo{author}{\bibfnamefont{M.}~\bibnamefont{Czakon}}, \bibnamefont{and}
  \bibinfo{author}{\bibfnamefont{T.}~\bibnamefont{Schutzmeier}},
  \bibinfo{journal}{Phys. Rev.} \textbf{\bibinfo{volume}{D74}},
  \bibinfo{pages}{074006} (\bibinfo{year}{2006}{\natexlab{a}}),
  \eprint{hep-ph/0605023}.

\bibitem[{\citenamefont{Maier et~al.}(2008{\natexlab{a}})\citenamefont{Maier,
  Maierhofer, and Marqaurd}}]{Maier:2008he}
\bibinfo{author}{\bibfnamefont{A.}~\bibnamefont{Maier}},
  \bibinfo{author}{\bibfnamefont{P.}~\bibnamefont{Maierhofer}},
  \bibnamefont{and} \bibinfo{author}{\bibfnamefont{P.}~\bibnamefont{Marqaurd}}
  (\bibinfo{year}{2008}{\natexlab{a}}), \eprint{0806.3405}.

\bibitem[{\citenamefont{Fleischer and Tarasov}(1994)}]{Fleischer:1994ef}
\bibinfo{author}{\bibfnamefont{J.}~\bibnamefont{Fleischer}} \bibnamefont{and}
  \bibinfo{author}{\bibfnamefont{O.~V.} \bibnamefont{Tarasov}},
  \bibinfo{journal}{Z. Phys.} \textbf{\bibinfo{volume}{C64}},
  \bibinfo{pages}{413} (\bibinfo{year}{1994}), \eprint{hep-ph/9403230}.

\bibitem[{\citenamefont{Broadhurst et~al.}(1993)\citenamefont{Broadhurst,
  Fleischer, and Tarasov}}]{Broadhurst:1993mw}
\bibinfo{author}{\bibfnamefont{D.~J.} \bibnamefont{Broadhurst}},
  \bibinfo{author}{\bibfnamefont{J.}~\bibnamefont{Fleischer}},
  \bibnamefont{and} \bibinfo{author}{\bibfnamefont{O.~V.}
  \bibnamefont{Tarasov}}, \bibinfo{journal}{Z. Phys.}
  \textbf{\bibinfo{volume}{C60}}, \bibinfo{pages}{287} (\bibinfo{year}{1993}),
  \eprint{hep-ph/9304303}.

\bibitem[{\citenamefont{Baikov and Broadhurst}(1995)}]{Baikov:1995ui}
\bibinfo{author}{\bibfnamefont{P.~A.} \bibnamefont{Baikov}} \bibnamefont{and}
  \bibinfo{author}{\bibfnamefont{D.~J.} \bibnamefont{Broadhurst}}
  (\bibinfo{year}{1995}), \eprint{hep-ph/9504398}.

\bibitem[{\citenamefont{Chetyrkin
  et~al.}(1997{\natexlab{a}})\citenamefont{Chetyrkin, Kuhn, and
  Steinhauser}}]{Chetyrkin:1997mb}
\bibinfo{author}{\bibfnamefont{K.~G.} \bibnamefont{Chetyrkin}},
  \bibinfo{author}{\bibfnamefont{J.~H.} \bibnamefont{Kuhn}}, \bibnamefont{and}
  \bibinfo{author}{\bibfnamefont{M.}~\bibnamefont{Steinhauser}},
  \bibinfo{journal}{Nucl. Phys.} \textbf{\bibinfo{volume}{B505}},
  \bibinfo{pages}{40} (\bibinfo{year}{1997}{\natexlab{a}}),
  \eprint{hep-ph/9705254}.

\bibitem[{\citenamefont{Hoang and Jamin}(2004)}]{Hoang:2004xm}
\bibinfo{author}{\bibfnamefont{A.~H.} \bibnamefont{Hoang}} \bibnamefont{and}
  \bibinfo{author}{\bibfnamefont{M.}~\bibnamefont{Jamin}},
  \bibinfo{journal}{Phys. Lett.} \textbf{\bibinfo{volume}{B594}},
  \bibinfo{pages}{127} (\bibinfo{year}{2004}), \eprint{hep-ph/0403083}.

\bibitem[{\citenamefont{Kuhn et~al.}(2007)\citenamefont{Kuhn, Steinhauser, and
  Sturm}}]{Kuhn:2007vp}
\bibinfo{author}{\bibfnamefont{J.~H.} \bibnamefont{Kuhn}},
  \bibinfo{author}{\bibfnamefont{M.}~\bibnamefont{Steinhauser}},
  \bibnamefont{and} \bibinfo{author}{\bibfnamefont{C.}~\bibnamefont{Sturm}},
  \bibinfo{journal}{Nucl. Phys.} \textbf{\bibinfo{volume}{B778}},
  \bibinfo{pages}{192} (\bibinfo{year}{2007}), \eprint{hep-ph/0702103}.

\bibitem[{\citenamefont{Hoang}(1999)}]{Hoang:1998uv}
\bibinfo{author}{\bibfnamefont{A.~H.} \bibnamefont{Hoang}},
  \bibinfo{journal}{Phys. Rev.} \textbf{\bibinfo{volume}{D59}},
  \bibinfo{pages}{014039} (\bibinfo{year}{1999}), \eprint{hep-ph/9803454}.

\bibitem[{\citenamefont{Hoang}(2000)}]{Hoang:1999ye}
\bibinfo{author}{\bibfnamefont{A.~H.} \bibnamefont{Hoang}},
  \bibinfo{journal}{Phys. Rev.} \textbf{\bibinfo{volume}{D61}},
  \bibinfo{pages}{034005} (\bibinfo{year}{2000}), \eprint{hep-ph/9905550}.

\bibitem[{\citenamefont{Czarnecki and Melnikov}(1998)}]{Czarnecki:1997vz}
\bibinfo{author}{\bibfnamefont{A.}~\bibnamefont{Czarnecki}} \bibnamefont{and}
  \bibinfo{author}{\bibfnamefont{K.}~\bibnamefont{Melnikov}},
  \bibinfo{journal}{Phys. Rev. lett.} \textbf{\bibinfo{volume}{80}},
  \bibinfo{pages}{2531} (\bibinfo{year}{1998}), \eprint{hep-ph/9712222}.

\bibitem[{\citenamefont{Hoang et~al.}(2002)\citenamefont{Hoang, Manohar,
  Stewart, and Teubner}}]{Hoang:2001mm}
\bibinfo{author}{\bibfnamefont{A.~H.} \bibnamefont{Hoang}},
  \bibinfo{author}{\bibfnamefont{A.~V.} \bibnamefont{Manohar}},
  \bibinfo{author}{\bibfnamefont{I.~W.} \bibnamefont{Stewart}},
  \bibnamefont{and} \bibinfo{author}{\bibfnamefont{T.}~\bibnamefont{Teubner}},
  \bibinfo{journal}{Phys. Rev.} \textbf{\bibinfo{volume}{D65}},
  \bibinfo{pages}{014014} (\bibinfo{year}{2002}), \eprint{hep-ph/0107144}.

\bibitem[{\citenamefont{Hoang et~al.}(2000)}]{Hoang:2000yr}
\bibinfo{author}{\bibfnamefont{A.~H.} \bibnamefont{Hoang}}
  \bibnamefont{et~al.}, \bibinfo{journal}{Eur. Phys. J. direct}
  \textbf{\bibinfo{volume}{C2}}, \bibinfo{pages}{1} (\bibinfo{year}{2000}),
  \eprint{hep-ph/0001286}.

\bibitem[{\citenamefont{Chetyrkin and Kuhn}(1994)}]{Chetyrkin:1994ex}
\bibinfo{author}{\bibfnamefont{K.~G.} \bibnamefont{Chetyrkin}}
  \bibnamefont{and} \bibinfo{author}{\bibfnamefont{J.~H.} \bibnamefont{Kuhn}},
  \bibinfo{journal}{Nucl. Phys.} \textbf{\bibinfo{volume}{B432}},
  \bibinfo{pages}{337} (\bibinfo{year}{1994}), \eprint{hep-ph/9406299}.

\bibitem[{\citenamefont{Chetyrkin
  et~al.}(1997{\natexlab{b}})\citenamefont{Chetyrkin, Harlander, Kuhn, and
  Steinhauser}}]{Chetyrkin:1997qi}
\bibinfo{author}{\bibfnamefont{K.~G.} \bibnamefont{Chetyrkin}},
  \bibinfo{author}{\bibfnamefont{R.}~\bibnamefont{Harlander}},
  \bibinfo{author}{\bibfnamefont{J.~H.} \bibnamefont{Kuhn}}, \bibnamefont{and}
  \bibinfo{author}{\bibfnamefont{M.}~\bibnamefont{Steinhauser}},
  \bibinfo{journal}{Nucl. Phys.} \textbf{\bibinfo{volume}{B503}},
  \bibinfo{pages}{339} (\bibinfo{year}{1997}{\natexlab{b}}),
  \eprint{hep-ph/9704222}.

\bibitem[{\citenamefont{Maier et~al.}(2008{\natexlab{b}})\citenamefont{Maier,
  Maierhofer, and Marquard}}]{Maier:2007yn}
\bibinfo{author}{\bibfnamefont{A.}~\bibnamefont{Maier}},
  \bibinfo{author}{\bibfnamefont{P.}~\bibnamefont{Maierhofer}},
  \bibnamefont{and} \bibinfo{author}{\bibfnamefont{P.}~\bibnamefont{Marquard}},
  \bibinfo{journal}{Nucl. Phys.} \textbf{\bibinfo{volume}{B797}},
  \bibinfo{pages}{218} (\bibinfo{year}{2008}{\natexlab{b}}),
  \eprint{0711.2636}.

\bibitem[{\citenamefont{Boughezal
  et~al.}(2006{\natexlab{b}})\citenamefont{Boughezal, Czakon, and
  Schutzmeier}}]{Boughezal:2006uu}
\bibinfo{author}{\bibfnamefont{R.}~\bibnamefont{Boughezal}},
  \bibinfo{author}{\bibfnamefont{M.}~\bibnamefont{Czakon}}, \bibnamefont{and}
  \bibinfo{author}{\bibfnamefont{T.}~\bibnamefont{Schutzmeier}},
  \bibinfo{journal}{Nucl. Phys. Proc. Suppl.} \textbf{\bibinfo{volume}{160}},
  \bibinfo{pages}{160} (\bibinfo{year}{2006}{\natexlab{b}}),
  \eprint{hep-ph/0607141}.

\bibitem[{\citenamefont{Baikov et~al.}(2004)\citenamefont{Baikov, Chetyrkin,
  and Kuhn}}]{Baikov:2004ku}
\bibinfo{author}{\bibfnamefont{P.~A.} \bibnamefont{Baikov}},
  \bibinfo{author}{\bibfnamefont{K.~G.} \bibnamefont{Chetyrkin}},
  \bibnamefont{and} \bibinfo{author}{\bibfnamefont{J.~H.} \bibnamefont{Kuhn}},
  \bibinfo{journal}{Nucl. Phys. Proc. Suppl.} \textbf{\bibinfo{volume}{135}},
  \bibinfo{pages}{243} (\bibinfo{year}{2004}).

\end{thebibliography}

\end{document}